\documentclass[prl,twocolumn,superscriptaddress,showpacs,amsmath,amstex,amssymb,citeautoscript]{revtex4-1}

\newcommand{\vect}[1]{\boldsymbol{#1}}

\usepackage[T1]{fontenc}
\usepackage[utf8]{inputenc}
\usepackage{lipsum}
\usepackage{amsmath}
\usepackage{amssymb}
\usepackage{bbm}
\usepackage{braket}
\usepackage{xcolor}
\usepackage{pifont}

\usepackage[mathscr]{euscript}
\usepackage[shortlabels]{enumitem}
\usepackage{graphicx}
\usepackage{lipsum}
\allowdisplaybreaks
\usepackage{float}
\usepackage{graphicx}
\usepackage{dsfont}
\usepackage{comment}
\usepackage[colorlinks=true]{hyperref}  
\hypersetup{
    bookmarks=true,         % show bookmarks bar?
    unicode=false,          % non-Latin characters 
    pdftoolbar=true,        % show Acrobat
    pdfmenubar=true,        % show Acrobat 
    pdffitwindow=false,     % window fit to page when opened
    pdfstartview={FitH},    % fits the width of the page to the window
    pdftitle={Moire phasons in twisted trilayer graphene},    % title
    pdfauthor={Samajdar et al.},     % author
    pdfsubject={},   % subject of the document
    pdfcreator={},   % creator of the document
    pdfproducer={}, % producer of the document
    pdfkeywords={} {} {}, % list of keywords
    pdfnewwindow=true,      % links in new window
    colorlinks=true,       % false: boxed links; true: colored links
    linkcolor=blue, %red,          % color of internal links (change box color with linkbordercolor)
    citecolor=blue,        % color of links to bibliography
    filecolor=magenta,      % color of file links
    urlcolor=blue           % color of external links
} 

\usepackage{lipsum}

\newcommand{\be}{\begin{equation}}
\newcommand{\ee}{\end{equation}}
\newcommand{\bea}{\begin{eqnarray}}
\newcommand{\eea}{\end{eqnarray}}

\newcommand{\Cdot}{\hspace*{-0.05cm}\cdot\hspace*{-0.05cm}}

\newcommand{\equref}[1]{Eq.~(\ref{#1})}

\newcommand{\figref}[1]{Fig.~\ref{#1}}

\newcommand{\diff}{\mathrm{d}}

\renewcommand{\approx}{\simeq}

\renewcommand{\vec}[1]{\boldsymbol{#1}}

\definecolor{wrongultramarine}{rgb}{1,0.5,0}

\usepackage{pdfpages}
\usepackage{pgffor}
\makeatletter
\AtBeginDocument{\let\LS@rot\@undefined}
\makeatother

\begin{document}

\title{Moir\'e phonons and impact of electronic symmetry breaking \\ in twisted trilayer graphene}

\author{Rhine Samajdar}
\affiliation{Department of Physics, Harvard University, Cambridge MA 02138, USA}

\author{Yanting Teng}
\affiliation{Department of Physics, Harvard University, Cambridge MA 02138, USA}

\author{Mathias S. Scheurer}
\affiliation{Institut f\"ur Theoretische Physik, Universit\"at Innsbruck, A-6020 Innsbruck, Austria}

\begin{abstract}
Twisted trilayer graphene is a particularly promising moir\'e superlattice system, due to its tunability, strong superconductivity, and complex electronic symmetry breaking.
Motivated by these properties, we study lattice relaxation and the long-wavelength phonon modes of this system.
We show that mirror-symmetric trilayer graphene hosts, aside from the conventional acoustic phonon modes, two classes of shear modes, which are even and odd under mirror reflection. The mirror-even modes are found to be gapless and equivalent to the ``phason'' modes of twisted bilayer graphene, with appropriately rescaled parameters. The modes odd under mirror symmetry have no analogue in twisted bilayer graphene and exhibit a finite gap, which we show is directly proportional to the degree of lattice relaxation.  
We also discuss the impact of mirror-symmetry breaking, which can be tuned by a displacement field or result from a stacking shift, and of rotational- as well as time-reversal-symmetry breaking, resulting from spontaneous electronic order. 
We demonstrate that this can induce finite angular momentum to the phonon branches. 
Our findings are important to the interpretation of recent experiments, concerning the origin of superconductivity and of linear-in-$T$ resistivity.
\end{abstract}

\maketitle

\begin{figure}[tb]
    \centering
    \includegraphics[width=\linewidth]{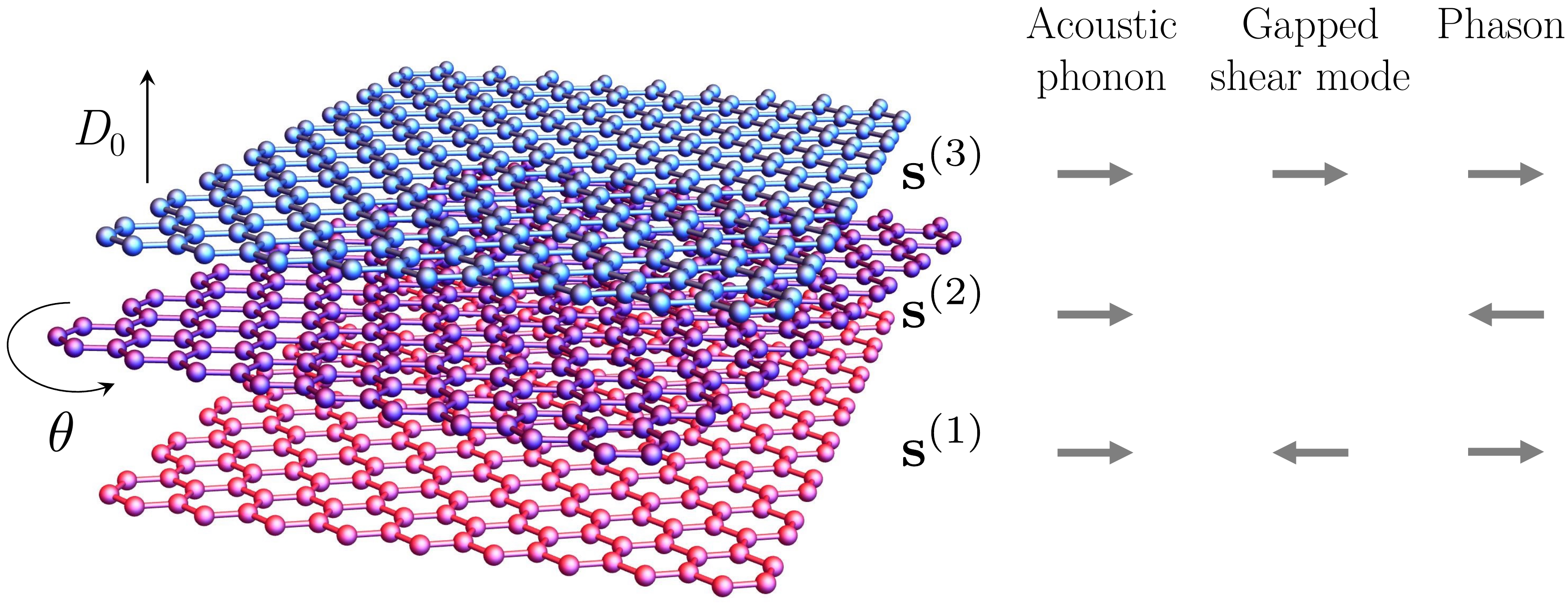}
    \caption{Illustration of TTG in a perpendicular displacement field $D_0$ and its low-energy phonon modes.}
    \label{LatticeAndModes}
\end{figure}

\textit{Introduction.---}Stacking and twisting different layers of graphene has emerged as a popular route to creating correlated superlattices over the last few years~\cite{macdonald2019bilayer,andrei2020graphene,kennes2020moir,balents2020superconductivity,STMReview,ZaletelJournalClub}. Besides the most well-studied system, twisted bilayer graphene (TBG), many other geometries have been explored both experimentally and theoretically. 
Among them, mirror-symmetric twisted trilayer graphene (TTG)~\cite{Park_2021,Hao_2021,2021arXiv210312083C,2021arXiv210912127K,2021arXiv210803338L,turkel2021twistons,OurDWPaper,DiodeExperiment,PhysRevLett.127.097001,TrilayerSC,PhysRevLett.127.217001,PhysRevB.104.174505,ChristosTrilayer,PhysRevB.104.115167,DiodeTheory,2021arXiv211111060L}, which consists of three graphene layers with alternating twist angles (see \figref{LatticeAndModes}) stands out: it is the first system that can be efficiently tuned with a perpendicular displacement field $D_0$ while exhibiting strong and reproducible superconductivity. Recently, another unique behavior was observed experimentally on decreasing the twist angle $\theta$ slightly below the magic angle~\cite{OurDWPaper,DiodeExperiment}: while the system still exhibits superconductivity with roughly the same critical temperature, the linear-in-temperature ($T$) resistivity seen at the magic angle disappears.
Motivated by these outstanding properties of TTG and the fact that both superconductivity and linear-in-$T$ resistivity \cite{Cao2020strange,Polshyn2019phonon} are considered to be linked to phonons in graphene moir\'e systems~\cite{Polshyn2019phonon,PhysRevB.99.165112,PhysRevB.99.140302,Ochoa,StepanovTuning,Liu1261,SaitoTuning}, here, we theoretically study thelong-wavelength and low-energy phonons of this system that are crucially determined by the moir\'e superlattice, and thus, depend on the twist angle.

Long-wavelength phonon modes have recently attracted significant attention~\cite{koshino,Ochoa,PhysRevB.104.064109,PhysRevLett.128.065901} in TBG. Besides the conventional acoustic phonons, the moir\'e lattice allows for gapless shear modes, often referred to as \textit{phasons}. Phason modes have a rich history in the study of quasicrystals \cite{Levine1985} and charge-density-wave materials \cite{PhasonCDW}. %More prosaically, 
In our case, they can be intuitively thought of as  the relative displacements of the two layers of TBG, which shift the moir\'e pattern and are, thus, gapless. However, a proper description~\cite{koshino,Ochoa} requires taking into account lattice relaxation, and the phasons then corresponds to the sliding of domain walls. 
Figure~\ref{LatticeAndModes} illustrates that TTG allows for three types of long-wavelengths phonon modes: on top of the regular acoustic phonons, where all three layers move in phase, there is another mirror-even set of modes, where the middle layer moves against the outer two layers. It is shown to be gapless and (modulo rescaling of parameters) exactly equivalent to the phasons of TBG. We further show that TTG also allows for mirror-odd shear modes, which have no analogue in TBG and are found to be gapped; the existence of a gap is intuitively understood by noting that a mirror-odd displacement of the layers does not just correspond to a translation or rotation of the moir\'e pattern, but rather to a nontrivial distortion of the superlattice. 

In this work, we not only investigate the twist-angle evolution of the phonon properties---which bear important consequences for the interpretation of experiments~\cite{OurDWPaper,DiodeExperiment}---but also the impact of reduced electronic symmetries on the phonon spectrum. This includes both the effects of an applied displacement field or stacking shifts, which break the mirror symmetry leading to an admixture of the gapless and gapped shear modes, and the consequences of spontaneous electronic symmetry breaking. Theoretical~\cite{ChristosTrilayer,PhysRevB.104.115167,DiodeTheory,2021arXiv211111060L} and experimental~\cite{Park_2021,Hao_2021,2021arXiv210312083C,2021arXiv210912127K,2021arXiv210803338L,turkel2021twistons,OurDWPaper,DiodeExperiment} studies of TTG indicate a variety of symmetry-broken electronic phases and studying their influence on the phonons is therefore crucial. We show how different electronic orders can induce finite angular momentum in the phonon branches.

% =================================================================
\textit{Formalism---}To begin, we first focus on the mirror-symmetric limit of TTG (i.e., in the absence of a displacement field) with no symmetry breaking in the electronic sector.
The total free energy is a sum of two pieces
$F$\,$=$\,$F^{}_{\textrm{el}}$\,$+$\,$F^{}_{\textrm{ad}}$, where 
$F^{}_{\textrm{el}}$ describes in-plane elastic distortions of the graphene layers and $F^{}_{\textrm{ad}}$ accounts for the interlayer adhesion energy. 
Labeling the layers from bottom to top by $l$\,$=$\,$1, 2, 3$, and their corresponding in-plane \cite{sm} displacements by the two-component fields $\mathbf{s}^{(l)}$, the first term can be written as~\cite{PhysRevB.65.235412}
\begin{align*}
F^{}_{\textrm{el}}=\sum_{l=1}^3\int \diff \vect{r}\,\left[\frac{\lambda}{2}\left(\boldsymbol{\nabla}\Cdot\mathbf{s}^{(l)}\right)^2+\frac{\mu}{4}\left(\partial^{}_i s_j^{(l)}+\partial^{}_j s_i^{(l)}\right)^2\right],
\end{align*}
where $\lambda \approx 3.25$~eV/\AA$^2$ and $\mu \approx 9.57$~eV/\AA$^2$ are the Lam\'e coefficients of graphene~\cite{PhysRevLett.102.046808,jung2015origin}. We briefly discuss the out-of-plane field component in supplement Sec.~IC~\cite{sm}.
The elastic theory is more naturally expressed in terms of the \textit{relative} displacements, $\vect{u} \equiv \mathbf{s}^{(3)}$\,$-$\,$ \mathbf{s}^{(1)}$, and $\vect{v}\equiv \mathbf{s}^{(3)}+\mathbf{s}^{(1)}-2\mathbf{s}^{(2)}$, which are odd and even under mirror reflections, respectively, and the total displacement, $\vect{w} \equiv \mathbf{s}^{(1)}+\mathbf{s}^{(2)}+\mathbf{s}^{(3)}$ (cf.~\figref{LatticeAndModes}). While changes in $\vect{u}$ and $\vect{v}$ correspond to shear modes, the mode $\vect{w}$ represents an in-phase displacement of all three graphene sheets. 
The adhesion energy $F^{}_{\textrm{ad}}$ is a functional of the relative displacement fields only, i.e., 
$
F_{\textrm{ad}}=\int \diff\vect{r}\,\mathcal{V}_{\textrm{ad}}\left[\vect{r},\vect{u}\left(\vect{r}\right), \vect{v}\left(\vect{r}\right)\right],
$
where $\mathcal{V}_{\textrm{ad}}$ represents the adhesion potential gluing the layers together; we expect $\mathcal{V}_{\textrm{ad}}$ to be well described as the sum of pairwise interlayer interactions between nearest-neighboring layers. Restricting its Fourier expansion to the smallest nonzero reciprocal lattice vectors $\vec{G}_\nu$, rotational symmetry implies that~\cite{sm}
\begin{align}\label{eq:potentialMAIN}
\mathcal{V}^{}_{\textrm{ad}}&= \sum_{l=1, 3}V^{}_{l}\sum_{\nu=1}^3 \cos\left[\frac{\mathbf{b}^{}_\nu}{2}\Cdot\left(\vect{v}+p^{}_{l}\vect{u}\right)-\mathbf{G}^{}_\nu\Cdot\vect{r} \right],
\end{align}
with $p_1$\,$=$\,$-1$, $p_3$\,$=$\,$+1$, where $\vec{b}_\nu$ are the reciprocal lattice vectors of a single graphene sheet;  mirror symmetry implies $V_1$\,$=$\,$V_3$\,$\equiv$\,$V$. Putting $F_{\textrm{el}}$ and $F_{\textrm{ad}}$ together, and solving the coupled Euler-Lagrange equations of motion for harmonic oscillations about the self-consistently determined equilibrium configurations $\{\vect{u}^{(0)}(\vect{r})$,\,$\vect{v}^{(0)}(\vect{r})\}$, we obtain the spectrum of lattice vibrations.

\begin{figure}[tb]
    \centering
    \includegraphics[width=\linewidth]{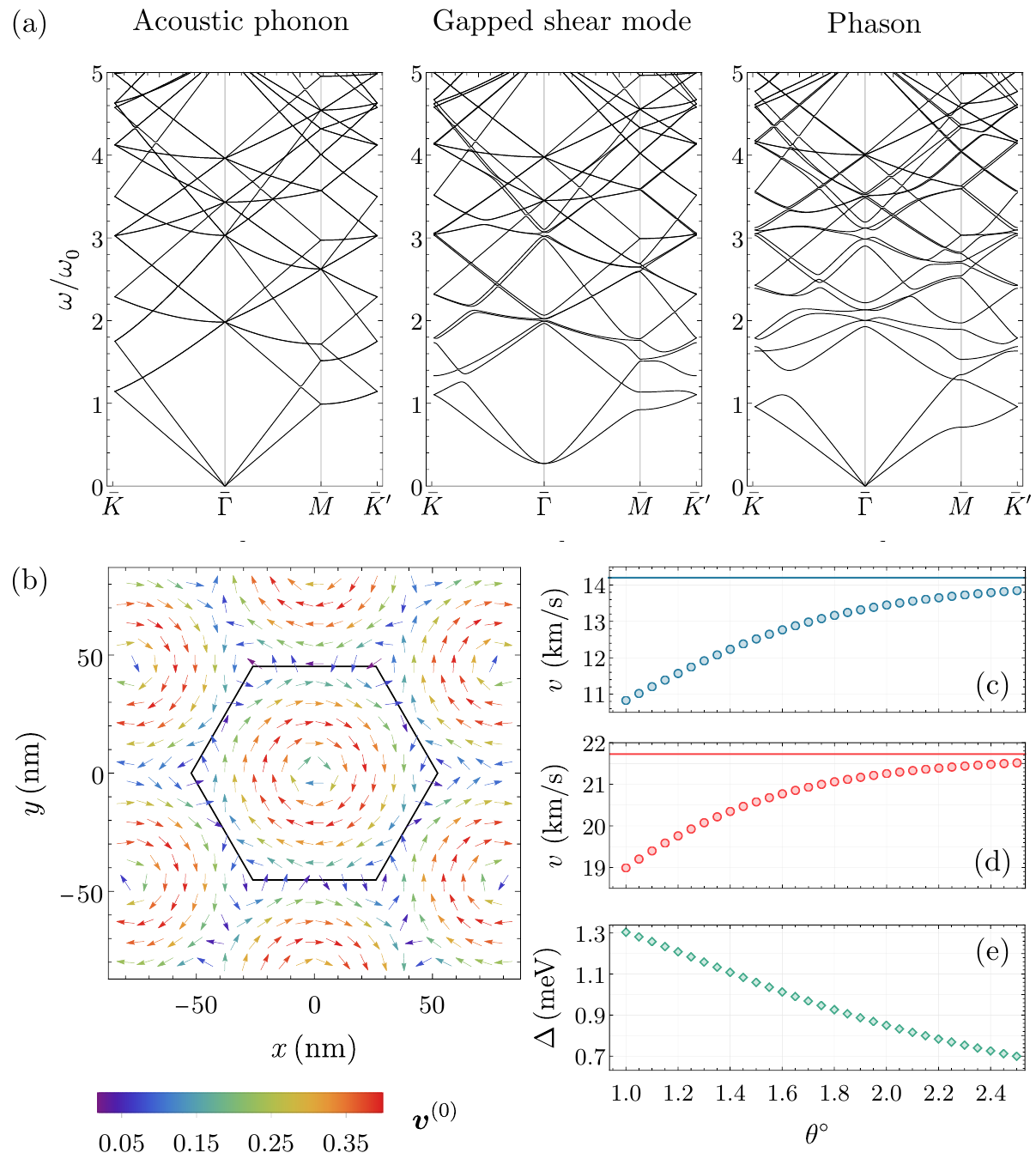}
    \caption{(a) Phonon spectra of TTG in the mirror-symmetric limit at the magic angle $\theta$\,$=$\,$1.56^\circ$. (b) Lattice relaxation texture for $\vect{v}^{(0)}$; the AA (AB/BA) stacking regions at the center (corners) of the moir\'e unit cell---demarcated by the solid hexagon---shrink (expand) under such relaxation. (c,d) Phason velocities for the lowest two branches; the solid lines mark the transverse ($v_{\textsc{ta}}$, blue) and longitudinal  ($v_{\textsc{la}}$, red) acoustic phonon velocities. (e) Gap of the mirror-odd shear mode as a function of the twist angle.}
    \label{fig:TTG}
\end{figure}

\textit{Mirror-symmetric limit.---}The relaxation of the moir\'e superlattice 
due to the interlayer couplings is captured by the displacement textures $\{\vect{u}^{(0)}$,\,$\vect{v}^{(0)}\}$.
We see that this atomic reconstruction leads to a nontrivial $\vect{v}^{(0)}$\,$\ne$\,$0$ [Fig.~\ref{fig:TTG}(b)], which illustrates that the lattice reorganizes itself to maximize (minimize) the regions of energetically (un)favorable AB/BA (AA) stackings in each bilayer. 
We find that this relaxation is equivalent to that of TBG, apart from a rescaling of parameters~\cite{sm}.
Furthermore, we obtain that $\vect{u}^{(0)}(\vect{r})$ is identically zero $\forall\,\, \vect{r}$. While spontaneous breaking of mirror symmetry by the lattice is possible in principle, $\vect{u}^{(0)}$\,$\ne$\,$0$ is not favorable energetically. This can be understood intuitively by noting that \textit{simultaneous} maximization of local AB/BA stacking regions in the top and bottom bilayers via $\vect{v}^{(0)}$\,$\ne$\,$0$ is most effective when the outer two layers are aligned, as seen in experiments \cite{2021arXiv210912127K,turkel2021twistons}.

The fully relaxed spectra of the three distinct classes of vibrational modes, arising from the displacements $\vect{u}$, $\vect{v}$, and $\vect{w}$, are arrayed in Fig.~\ref{fig:TTG}(a) in units of $\omega_0= (2\pi/L_M)\sqrt{\lambda/\rho}$, where $\rho$ is the mass density and $L_M$ is the moir\'e lattice constant. Physically, the acoustic phonon modes, $\vect{w}(\vect{q})$, represent the in-phase vibrations of all three layers. 
On the other hand, the gapless \textit{phason} modes, $\vect{v}(\vect{q})$, correspond to the sliding motion of the domain walls. As shown in Sec.~IB of the SM~\cite{sm}, the phason mode for TTG is equivalent to the one of TBG modulo rescaling of the adhesion potential $V$\,$\rightarrow$\,$ 2V/3$.
The phasons can be thought of as acoustic modes of the emergent moir\'e superlattice, and the soft nature of this lattice can be seen from the velocities in Fig.~\ref{fig:TTG}(c) and (d). Unlike the acoustic mode velocities, which, within the harmonic approximation, are just constants given by $v_{\textsc{la}}$\,$=$\,$\sqrt{(\lambda+2\mu)/\rho}$ and $v_{\textsc{ta}}$\,$=$\,$\sqrt{\mu/\rho}$, for longitudinal and transverse phonons, respectively, we see that the velocities of the low-frequency phason modes are extremely sensitive to twisting and can thus be used as an indirect probe to infer the twist angle. The existence of these soft phason modes is also expected to modify (and imprint signatures in) various experimental observables, including the low-temperature specific heat~\cite{nika2014specific}, thermal conductivity~\cite{li2014thermal}, and frictional properties such as superlubricity~\cite{maity2020phonons,PhysRevB.93.201404,PhysRevB.94.045401,PhysRevB.99.054103}.

Finally, there also exists a gapped shear mode, $\vect{u}(\vect{q})$, which is unique to TTG. This mode corresponds to a distortion of the moir\'e lattice and is thus massive but the  $C_3$ symmetry guarantees that the lowest two branches (the two polarizations) have the same mass.
In Fig.~\ref{fig:TTG}(e), we see that the gap $\Delta$ decreases monotonically with increasing twist angles and this behavior of the gap can be understood from the expression
\begin{equation}
    \Delta^2 = \frac{3 V}{2\rho}|\vec{b}|^2 \alpha; \,\,\, \alpha=\left\langle-\cos\big(\vec{b}^{}_\nu\vec{v}^{(0)}/2-\vec{G}^{}_\nu\vec{r}\big) \right\rangle_\Omega, \label{ExpressionForMass}
\end{equation}
obtained from perturbation theory (see Eq.~(S32)~\cite{sm}).
Here, $|\vec{b}|$\,$\equiv$\,$|\vec{b}_\nu|$, and $\braket{\dots}_\Omega$ denotes the spatial average over the system. Without any relaxation, i.e., $\vec{v}_0(\vec{r})$\,$=$\,$0$, we have $\alpha$\,$=$\,$0$, while relaxation will result in $\alpha$\,$>$\,$0$ to optimize the adhesion potential. This is also expected since, in the absence of relaxation, the top and bottom layers can be moved independently without any energetic cost in our elastic theory. Therefore, the gap of the mirror-odd shear mode---a \textit{thermodynamically observable} quantity---is directly proportional to the dimensionless measure $\alpha$ of lattice relaxation. Furthermore, we now also immediately understand that, for smaller twist angles, the impact of the relaxation will be stronger such that $\alpha$, and hence, $\Delta$, increases, in accordance with \figref{fig:TTG}(e). Given that the gap varies from $10.5$ to $5.6$~cm$^{-1}$ over this range of $\theta$, the mirror-odd shear modes can be directly probed by Brillouin-Mandelstam spectroscopy \cite{speziale2014brillouin,kargar2016direct}.

% =================================================================

\begin{figure}[tb]
    \centering
    \includegraphics[width=\linewidth]{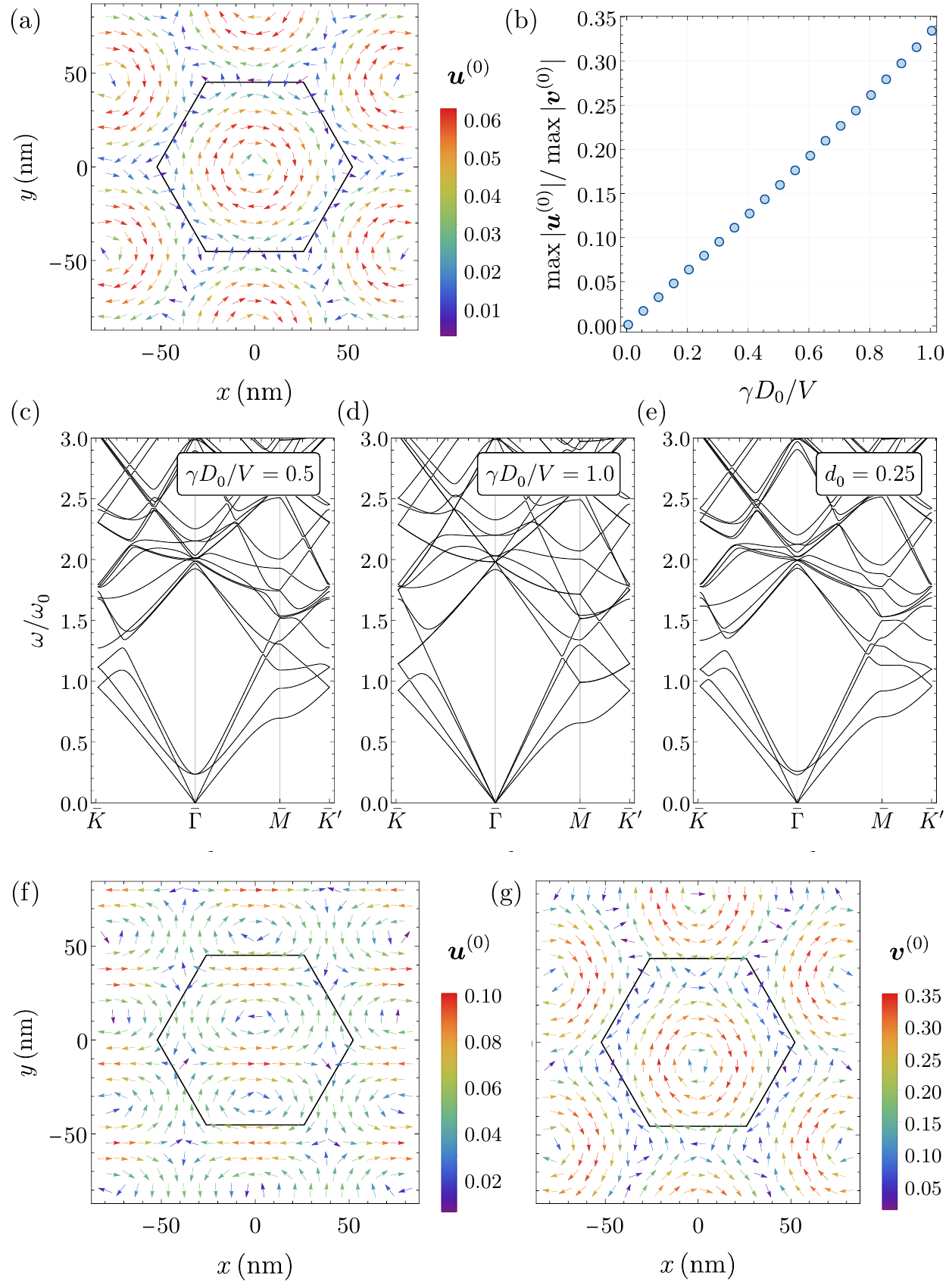}
    \caption{(a) Mirror-odd lattice relaxation $\vect{u}^{(0)}$, and (b) its strength relative to the mirror-even $\vect{v}^{(0)}$ for $\theta$\,$=$\,$1.56^\circ$ TTG in the presence of a displacement field. (c, d) Spectra of shear modes as the field is varied. (e) The phonon spectrum for the shear modes when the mirror-symmetry breaking is induced by a lateral stacking shift of magnitude $d_0$\,$=$\,$0.25$; for clarity, we omit the acoustic phonons in (c-e). (f) and (g) show the corresponding relaxation textures in this case. }
    \label{ResultsWithDisplacementField}
\end{figure}

\textit{Mirror-symmetry breaking.---}One particularly interesting aspect of TTG is that an electric displacement field $D_0$ can be applied perpendicular to the graphene layers (see \figref{LatticeAndModes}), which breaks the mirror symmetry. $D_0$ is known to strongly affect the electronic degrees of freedom as seen in experiments~\cite{Park_2021,Hao_2021,2021arXiv210312083C,2021arXiv210912127K,2021arXiv210803338L,turkel2021twistons,OurDWPaper,DiodeExperiment} and thus, is expected to also modify the phononic properties. In our elastic theory, we model this phenomenologically by allowing for the adhesion potential strength to differ between the bottom and top bilayers, taking $V_{1,3} = V \mp \gamma D_0$ in \equref{eq:potentialMAIN}. 

The broken mirror symmetry has two crucial consequences. Firstly, while $\vect{v}^{(0)}$ continues to resemble the previously found profile in \figref{fig:TTG}(b), lattice relaxation now also occurs in the mirror-odd sector, i.e., $\vec{u}^{(0)}$\,$\neq$\,$0$. This can be seen in Figs.~\ref{ResultsWithDisplacementField}(a,b) where we plot the texture $\vec{u}^{(0)}(\vec{r})$, which closely follows that of $\vec{v}^{(0)}$, and how its strength evolves approximately linearly with $\gamma D_0$. Specifically, we see that $\vec{u}^{(0)}$\,$\rightarrow$\,$\vec{v}^{(0)}/3$ as the extreme limit $\gamma D_0/V$\,$=$\,$1$ is approached; 
this corresponds to the absence of relaxation in the top layer, which becomes completely decoupled in this case. 
While slightly unphysical, this limit helps us understand qualitatively the behavior of the second key modification---the change of the shear mode spectra. As can be seen in \figref{ResultsWithDisplacementField}(c), the gapless phason mode at $D_0=0$ stays gapless for finite $D_0$ while the gap of the originally mirror-odd mode decreases. This is because the spectrum must approach that of the acoustic phonon of a single graphene layer and the phason of TBG (with doubled adhesion potential) in the abovementioned limit of $\gamma D_0/V$\,$=$\,$1$.

Another route to breaking mirror symmetry is via lateral stacking shifts~\cite{PhysRevB.104.035139}, which naturally arise in experimental samples. For concreteness, let us consider a case in which the topmost layer is displaced from the original ``A-twist-A'' stacking by a vector $\vect{d}$\,$=$\,$d_0\, \vect{a}/2$; 
here, $\vect{a}$ is chosen such that $d_0$\,$=$\,$1$ corresponds to ``A-twist-B'' stacking, which we find to be structurally unstable, as signaled by imaginary phonon frequencies. Interestingly, however, for $d_0$\,$\le$\,$0.5$, we discover \textit{metastable} configurations that are fundamentally  distinct from our previous ones. 
This is most clearly seen from the lattice relaxation textures in \figref{ResultsWithDisplacementField}(f,g): since $\vect{u}^{(0)}$ measures the static shift between the outer layers, the fact that it is a nonconstant vector (i.e., $\vect{u}^{(0)}$\,$\ne$\,$-\vect{d}$\,\,$\forall$\,\,$\vect{r}$) implies that the system does not just relax back to the earlier mirror-symmetric configuration but instead finds a different local minimum, oscillations about which yield the phonon spectrum in \figref{ResultsWithDisplacementField}(e). The broken threefold rotational symmetry now lifts the prior degeneracy of the mirror-odd shear modes at the $\Gamma$ point.

% =================================================================
\textit{Spontaneous electronic symmetry breaking.---}Besides explicit symmetry breaking via external fields, TTG also exhibits a variety of electronic phases with spontaneously broken symmetry, as indicated by experiments~\cite{Park_2021,Hao_2021,2021arXiv210312083C,2021arXiv210912127K,2021arXiv210803338L,turkel2021twistons,OurDWPaper,DiodeExperiment}; some of these states coexist with superconductivity~\cite{Park_2021,Hao_2021,2021arXiv210803338L,DiodeExperiment} and appear in the same regime of the phase diagram as the linear-in-$T$ resistivity. As such, it is important to analyze the consequences of these electronic orders for the phonons. 
The combination $C_{2z}\Theta$ of time-reversal ($\Theta$) and twofold-rotation symmetry ($C_{2z}$) is not only relevant to the stability of the electronic Dirac cones of the system, but also for the phonon modes: the phononic angular momentum~\cite{PhysRevLett.112.085503},
$
		L^z_{\vec{q}}$\,$=$\,$\int\diff\vec{r} \sum_{l=1}^{3} (\delta \mathbf{s}^{(l)}_{\vec{q}}\times \delta \dot{\mathbf{s}}^{(l)}_{\vec{q}} )_z, 
$
is constrained by $\Theta$ ($C_{2z}$) to obey $L^z_{\vec{q}}$\,$=$\,$-L^z_{-\vec{q}}$ ($L^z_{\vec{q}}$\,$=$\,$L^z_{-\vec{q}}$) and thus, vanishes if $C_{2z}\Theta$ is a symmetry \footnote{More generally, it vanishes if there is a valley-U(1) or spin rotation that, when combined with $C_{2z}\Theta$, is a symmetry.}. Recent theory~\cite{ChristosTrilayer} finds the emergence of sublattice-polarized phases at finite $D_0$, which break $C_{2z}\Theta$, gapping out all Dirac cones, in consistency with the more resistive behavior seen experimentally~\cite{Park_2021,Hao_2021} in this regime. Whether this proceeds via breaking of $C_{2z}$ or $\Theta$ depends on details of the exact interactions present and can be thought of as the spontaneous emergence of loop currents on the moir\'e scale with opposite or same chirality in the two valleys, respectively.

\begin{figure}[tb]
    \centering
    \includegraphics[width=\linewidth]{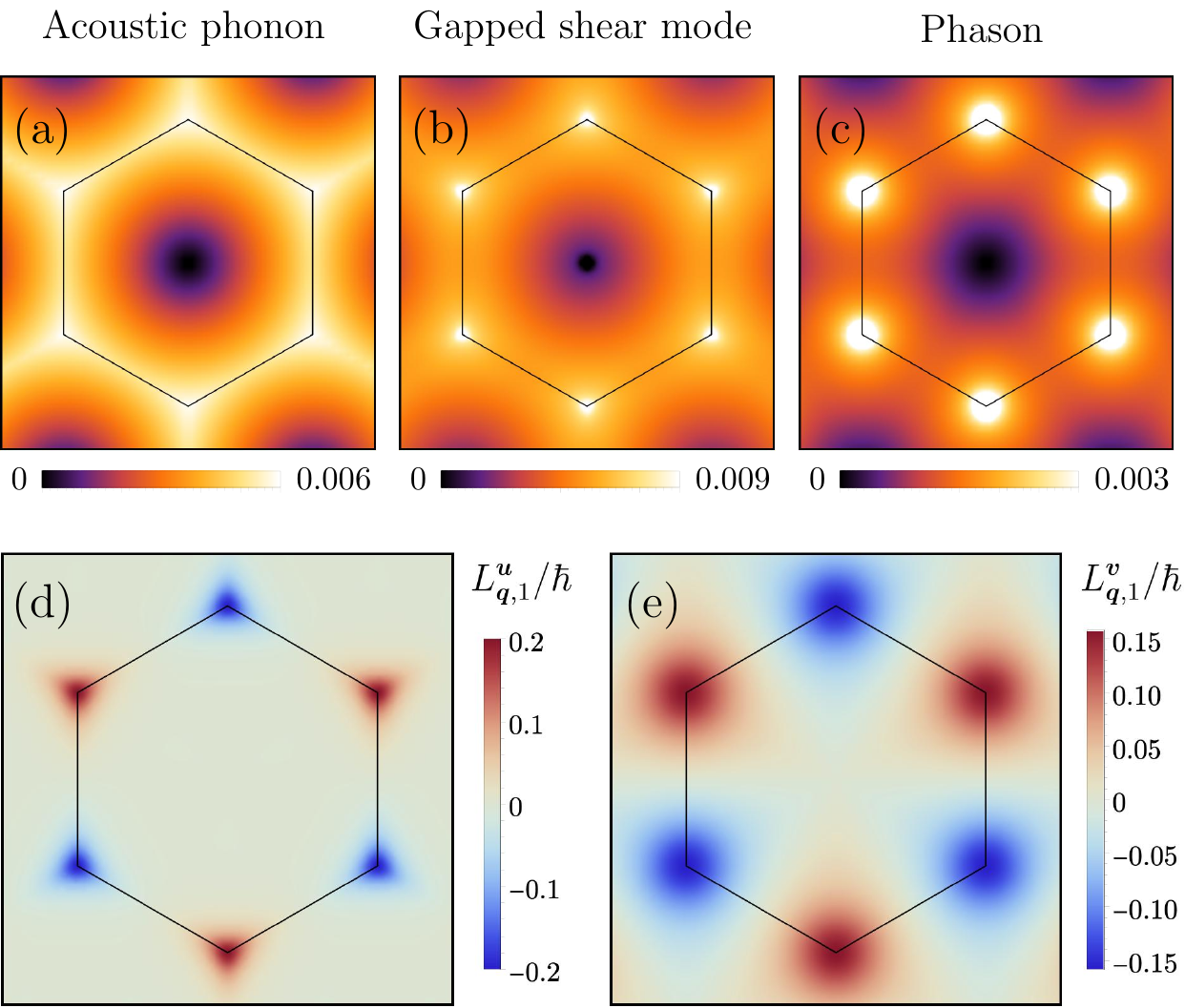}
    \caption{Angular momentum of the lowest band of the (a) acoustic phonon, (b) gapped shear, and (c) phason modes in the presence of broken time-reversal symmetry at $\theta$\,$=$\,$1.56^\circ$, using a phenomenological value of $\eta/\sqrt{\rho}=1\, (\mathrm{eV})^{1/2}$ (see Fig.~S4~\cite{sm} for the variation with $\eta$). The lower panel shows the angular momentum for the lowest band of the (d) gapped shear mode, and (e) phason when $C_{2z}$ is broken for a value of $\phi$\,$=$\,$\pi/6$ (cf.~Fig.~S5~\cite{sm} for variation in $\phi$). In this case, the angular momentum of the acoustic phonon is identically $0$.}
    \label{ResultsWithReducedElectronicSymmetries}
\end{figure}

To begin with the latter, broken time-reversal symmetry induces a Hall viscosity term in the elastic theory~\cite{PhysRevLett.75.697,HallVisc2}, which---owing to $C_{6z}$ rotation symmetry---can be parameterized (Sec.~SIIIB~\cite{sm}) by a single real number, $\eta$, as 
$F_{\eta}$\,$=$\,$\eta \int\diff\vec{r}\sum_{l}  ([\partial^2_j \delta \mathbf{s}^{(l)}]$\,$\times$\,$\delta \dot{\mathbf{s}}^{(l)})_z$.
Here, $\eta$ has to scale as $\eta$\,$\sim$\,$g^2 M$ for small electron-phonon coupling strength $g$ and magnitude $M$ of the sublattice polarization~\cite{2021PhRvB.104c5103Z}. As shown in \figref{ResultsWithReducedElectronicSymmetries}(a-c), this induces a finite angular momentum in all low-energy modes, the overall scale of which increases with $\eta$~\cite{sm}.
Most importantly, by virtue of resulting from broken $\Theta$ rather than $C_{2z}$ symmetry, the integral of $L^z_{\vec{q}}$ over the Brillouin zone of a given band does not vanish, which crucially differs from previous discussions of angular momentum bands in moir\'e systems~\cite{DiChiral,LischnerChiral}. Interestingly, the contributions from the different modes have the same sign, as opposed to acoustic and optical phonons in regular crystals~\cite{PhysRevLett.112.085503}. Consequently, the phononic system exhibits a finite ground-state angular momentum, which has to decay at sufficiently large temperature due to the Bohr-van Leeuwen theorem. Fingerprints of the angular momenta in the phononic bands could potentially even be probed experimentally via the 
Einstein-de Haas effect~\cite{PhysRevLett.112.085503} or the phonon thermal Hall effect~\cite{QinThermalHall}, which has attracted much attention recently in the context of cuprate superconductors \cite{Grissonnanche_2020} and Kitaev materials \cite{PhysRevX.12.021025}. For completeness, we also studied~\cite{sm} sublattice polarization which breaks $C_{2z}$ instead of $\Theta$; practically, this corresponds to adding a phase $\phi$ to each cosine of $\mathcal{V}^{}_{\textrm{ad}}$ in Eq.~\eqref{eq:potentialMAIN}. In this case, $L^z_{\vec{q}}$ is odd in $\vec{q}$, as clearly seen in \figref{ResultsWithReducedElectronicSymmetries}(d-e), and the net angular momentum vanishes at any temperature.

\textit{Discussion.---}Making the natural~\cite{2021arXiv210803338L} assumption that electron-phonon coupling is an important driving force of superconductivity in TBG and TTG, our results provide a natural explanation for why both systems show superconductivity with comparable $T_c$. As we have shown, the phason modes are equivalent in the two systems, modulo an $\mathcal{O}(1)$ rescaling of parameters, and we expect the additional mirror-odd mode to only provide a subleading enhancement of $T_c$ for vanishing $D_0$, since it couples the flat (TBG-like) bands to the highly dispersive (graphene-like) bands of TTG. When turning on a finite $D_0$, the two sectors mix and we expect the gapped shear mode to become more relevant; 
this might play an important role in the observed enhancement~\cite{Park_2021} of superconductivity for small $D_0$. 
In this picture, our results are also consistent with recent experiments, where the transition temperature was found to be approximately the same at the magic angle ($\theta$\,$\approx$\,$1.5^\circ$) and in the ``small-twist-angle  regime'' ($\theta$\,$\approx $\,$1.3^\circ$)~\cite{2021arXiv211207841L}, since we find that the phonon properties change by only a small amount in this range of $\theta$, see \figref{fig:TTG}(c--e).

If electron-phonon coupling is also responsible for the linear-in-$T$ resistivity, $\varrho\propto T$, which is considered to be a plausible scenario~\cite{Polshyn2019phonon,PhysRevB.99.165112,PhysRevB.99.140302,Ochoa}, the observed suppression \cite{OurDWPaper} of it around $\theta$\,$=$\,$1.3^\circ$ will have to be due to at least one of the following reasons: 
(\textit{i}) the temperature scale below which phonons do not give rise to a linear-in-$T$ contribution increases significantly. This temperature scale is an $\mathcal{O}(1)$ fraction~\cite{PhysRevB.77.115449} of the Bloch-Gr\"uneisen-like temperature scale $T_{\textsc{bg}}$\,$=$\,$\omega(\vec{k}^*)$ where $\vec{k}^*$ is the characteristic momentum transfer required to change the direction of the electronic group velocity significantly. Choosing  $\vec{k}^*$ to be half the vector connecting the $\Gamma$ and $K$ point as an example, we find $T_{\textsc{bg}} \approx 25\,\text{K}$ and $30\,\text{K}$ for the mirror-even and mirror odd modes at $\theta$\,$=$\,$1.5^\circ$. These scales even decrease further, to $18\,\text{K}$ and $24\,\text{K}$, respectively, when reducing the angle to $\theta$\,$=$\,$1.3^\circ$. 
So, we are left with possibility (\textit{ii}) that the magnitude of the phonon-induced contribution to $\varrho$ decreases rapidly with $\theta$. Focusing on the gapless phason mode, the slope $\diff\varrho/\diff T$ is proportional to~\cite{Ochoa,PhysRevB.99.165112} $|g|^2/(v^2_{\textsc{f}} \rho_m v^2_{\text{ph}})$, where $g$ is the electron-phonon coupling matrix element, $v_{\textsc{f}}$\,($v_{\text{ph}}$) the Fermi (phonon) velocity, and $\rho_m \propto \sin(\theta/2)$ the soliton-network mass density~\cite{Ochoa}. From our phonon spectra, we find $\rho_m v_{\text{ph}}^2$ decreases by about $20\%$ from $\theta = 1.5^\circ$ to $1.3^\circ$, corresponding to an increased contribution to $\varrho$. Furthermore, the  approximately identical phonon spectra at these two angles imply that the resulting superconducting $T_c$ should be primarily determined by $|g|^2/v_{\textsc{f}}$, since the density of states at a fixed filling fraction scales as $1/v_{\textsc{f}}$. 
Experimentally, $T_c$ is seen~\cite{OurDWPaper} to be about the same for the two angles, so the only way to explain the absence of linear-in-$T$ behavior at small $\theta$ is if $v_{\textsc{f}}$ increases significantly from $\theta$\,$=$\,$1.5^\circ$ to $1.3^\circ$; this, however, is not plausible either as the measured~\cite{OurDWPaper} bandwidth is even smaller in the small twist-angle regime.
Consequently, the recent data of Ref.~\onlinecite{OurDWPaper} is not consistent with a picture based on phonons alone and points towards another origin. This is not only in accordance with a very recent low-temperature study on TBG~\cite{Jaoui2021quantum} but is also reminiscent of transport behavior in the ``strange metal'' phase of the cuprate superconductors \cite{sachdev2016novel}.

\vspace{1em}

\begin{acknowledgments}
We thank Maine Christos, Jia Li, Daniel Parker, and Yunchao Zhang for useful discussions. R.S. and Y.T. acknowledge funding from the National Science Foundation under grant DMR-2002850. M.S.S.~acknowledges funding from the European Union (ERC-2021-STG, Project 101040651---SuperCorr). Views and opinions expressed are however those of the authors only and do not necessarily reflect those of the European Union or the European Research Council. Neither the European Union nor the granting authority can be held responsible for them.\\
\textit{Note added:} After the completion of this work, we became aware of a recent study discussing the symmetry origins of lattice vibrational modes in TTG~\cite{eslam2022}.
\end{acknowledgments}

% \DeclareFieldFormat*{title}{\mkbibquote{#1\isdot}} 
\bibliography{draft_Refs.bib}

\newpage


\section*{Supplementary material}

# Supplemental Material for "Moiré phonons and impact of electronic symmetry breaking in twisted trilayer graphene"

Rhine Samajdar,[1] Yanting Teng,[1] and Mathias S. Scheurer[2]

[1]*Department of Physics, Harvard University, Cambridge MA 02138, USA*
[2]*Institut für Theoretische Physik, Universität Innsbruck, A-6020 Innsbruck, Austria*

## SI. ELASTIC THEORY OF MOIRÉ PHONONS

In this section, we derive the Euler-Lagrange equations of motion that govern the lattice vibrations on the moiré scale, closely following the approach outlined in Refs. [1] and [2]. We will begin by developing the formalism for the general case where all symmetries of TTG are preserved, and extensions to more specific symmetry-breaking instances are described in subsequent sections.

The elastic free energy consists of two parts:

$$F = F_{\rm el} + F_{\rm ad};  \qquad (S1)$$

the first term describes in-plane elastic distortions of the graphene layers, whereas the second accounts for the interlayer adhesion energy. Labeling the layers from bottom to top by $l = 1, 2, 3$, we have

$$F_{\rm el} = \sum_{l=1}^{3} \int d\boldsymbol{r} \left[ \frac{\lambda_l}{2} \left( \boldsymbol{\nabla} \cdot \boldsymbol{s}^{(l)} \right)^2 + \frac{\mu_l}{4} \left( \partial_i s_j^{(l)} + \partial_j s_i^{(l)} \right)^2 \right], \qquad (S2)$$

where $\lambda_l$ and $\mu_l$ are the Lamé coefficients of the $l^{\rm th}$ graphene layer. While mirror symmetry technically only implies $\lambda_1 = \lambda_3$ and $\mu_1 = \mu_3$ (which, in general, can differ from $\lambda_2$ and $\mu_2$), we choose $\lambda_l = \lambda$ and $\mu_l = \mu$ since we expect the impact of the neighboring graphene layers on the *intra*layer elastic constants to be weak. For the same reason, henceforth, we use the values, $\lambda \simeq 3.25$ eV/Å$^2$ and $\mu \simeq 9.57$ eV/Å$^2$, which are the elastic constants of monolayer graphene [3, 4]. In Eq. (S2), the two-component fields $\boldsymbol{s}^{(l)}(\boldsymbol{r}, t)$ denote the in-plane displacements of the unit cells in layer $l$ with respect to their equilibrium positions in the absence of any interlayer forces. It is now useful to recast Eq. (S2) in terms of the relative displacements, $\boldsymbol{u} \equiv \boldsymbol{s}^{(3)} - \boldsymbol{s}^{(1)}$ and $\boldsymbol{v} \equiv \boldsymbol{s}^{(3)} + \boldsymbol{s}^{(1)} - 2\boldsymbol{s}^{(2)}$, which are odd and even under mirror reflections, respectively, and the total displacement, $\boldsymbol{w} \equiv \boldsymbol{s}^{(1)} + \boldsymbol{s}^{(2)} + \boldsymbol{s}^{(3)}$. Using these new variables, the in-plane free energy assumes the form

$$F_{\rm el} = \int d\boldsymbol{r} \left[ \frac{\lambda}{12} \left\{ 3 \left( \boldsymbol{\nabla} \cdot \boldsymbol{u} \right)^2 + \left( \boldsymbol{\nabla} \cdot \boldsymbol{v} \right)^2 + 2 \left( \boldsymbol{\nabla} \cdot \boldsymbol{w} \right)^2 \right\} \right. \\ \left. + \frac{\mu}{6} \left( 3 u_{ij}^2 + v_{ij}^2 + 2 w_{ij}^2 \right) \right], \qquad (S3)$$

where $u_{ij} \equiv (\partial_i u_j + \partial_j u_i)/2$, and likewise for $v_{ij}$, $w_{ij}$.

Next, we turn to the second term in Eq. (S1), namely, the adhesion energy, which describes energetic variations between different stacking configurations of the layers resulting from how the individual atoms are positioned. This energy should be a functional of the relative displacement fields only, i.e.,

$$F_{\rm ad} = \int d\boldsymbol{r} \, \mathcal{V}_{\rm ad} \left[ \boldsymbol{r}, \boldsymbol{u}(\boldsymbol{r}), \boldsymbol{v}(\boldsymbol{r}) \right], \qquad (S4)$$

where $\mathcal{V}_{\rm ad}$ is called the adhesion potential. In order to derive $\mathcal{V}_{\rm ad}$, we first characterize the local structure of TTG. Let us consider an *unrotated* trilayer heterostructure in which the carbon atoms of the three layers lie on top of each other. Upon twisting by angles $\{-\theta/2, +\theta/2, -\theta/2\}$, an atom on the second layer originally located at site $\boldsymbol{r}_0$—right above (below) the corresponding atom on the first (third) layer—shifts to a new position $\boldsymbol{r} = \mathcal{R}(\theta) \, \boldsymbol{r}_0$. We thus define the interlayer sliding vector [5], which measures the in-plane position of an atom in the second layer with respect to its counterparts in the other two layers. Assuming a rigid and static moiré lattice for TTG, this vector is the same for measurements relative to either the top or the bottom layer and is given by

$$\boldsymbol{\Delta}_0(\boldsymbol{r}) = \boldsymbol{r} - \boldsymbol{r}_0 = [1 - \mathcal{R}(-\theta)] \, \boldsymbol{r}. \qquad (S5)$$

However, accounting for the in-plane lattice vibrations specified by the displacements $\boldsymbol{s}^{(l)}(\boldsymbol{r}, t)$, we obtain *two* time-dependent sliding vectors

$$\begin{aligned} \boldsymbol{\Delta}_1(\boldsymbol{r}, t) &= \boldsymbol{\Delta}_0(\boldsymbol{r}) + \boldsymbol{s}^{(2)}(\boldsymbol{r}, t) - \boldsymbol{s}^{(1)}(\boldsymbol{r}, t), \\ \boldsymbol{\Delta}_3(\boldsymbol{r}, t) &= \boldsymbol{\Delta}_0(\boldsymbol{r}) + \boldsymbol{s}^{(2)}(\boldsymbol{r}, t) - \boldsymbol{s}^{(3)}(\boldsymbol{r}, t), \end{aligned} \qquad (S6)$$

describing the shifts within the bilayers formed by layers $l = 1, 2$, and $l = 2, 3$, respectively. Intuitively, this means that when the moiré lattice constant $L_{\rm M} \gg a$, the local lattice structure for the former (latter) bilayer at position $\boldsymbol{r}$ resembles that of unrotated bilayer graphene with the sheets relatively shifted by $\boldsymbol{\Delta}_1(\boldsymbol{r}, t)$ $(\boldsymbol{\Delta}_3(\boldsymbol{r}, t))$ [5].

The adhesion potential $\mathcal{V}_{\rm ad}$ is periodic in the fields $\boldsymbol{\Delta}_l$ $(l = 1, 3)$ with the lattice period of graphene [5]. Temporarily suppressing the time dependences, the simplest form of such potential compatible with sixfold symmetry is to keep the first six Fourier harmonics (*first star approximation*),

$$\mathcal{V}_{\rm ad} \left[ \boldsymbol{r}, \boldsymbol{\Delta}_1(\boldsymbol{r}), \boldsymbol{\Delta}_2(\boldsymbol{r}) \right] = \sum_{l=1,3} V_l \left[ \sum_{\nu=1}^{3} \cos \left( \boldsymbol{b}_\nu \cdot \boldsymbol{\Delta}_l \right) \right], \qquad (S7)$$

where mirror symmetry implies $V_1 = V_3 = V$ and $\boldsymbol{b}_\nu$ denotes the primitive vector of the reciprocal lattice of layer

$\nu$; we set $V = 90$ meV/nm² [2] based on first-principles calculations by Carr *et al.* [6]. We note, however, that there is significant variation in the parameters reported in the literature, with some studies—using several competing density-functional-theory methods—estimating a range of $V \sim 39$–$104$ meV/nm² [7, 8], whereas others suggest a larger value of $V \sim 320$ meV/nm² [1, 9, 10]. The function (S7) results in a maximum value for $\mathcal{V}_{ad}$ of $6V$ at AAA stackings $[\boldsymbol{\Delta}_l = 0]$ and a minimum value of $-3V$ at ABA or BAB stackings $[\boldsymbol{\Delta}_l = (a/2, a/(2\sqrt{3}))]$. From Eq. (S7), it is also easy to see that the energetically favorable atomic arrangement for TTG is a locally mirror symmetric AtA ("A-twist-A") trilayer configuration comprised of AAA, ABA, and BAB stacking sites [11], so that the potential energies for each bilayer can be optimized simultaneously. Plugging Eq. (S6) into Eq. (S7) and noting that $\mathbf{b}_\nu \cdot \boldsymbol{\Delta}_0(\boldsymbol{r}) = \mathbf{G}_\nu \cdot \boldsymbol{r}$, where $\mathbf{G}_\nu$ is a moiré reciprocal lattice vector, we find

$$\mathcal{V}_{ad}\left[\boldsymbol{r}, \mathbf{s}^{(l)}(\boldsymbol{r})\right] = V \sum_{\nu=1}^{3} \left\{ \cos\left[\mathbf{G}_\nu \cdot \boldsymbol{r} + \mathbf{b}_\nu \cdot \left(\mathbf{s}^{(2)} - \mathbf{s}^{(1)}\right)\right] \right.$$
$$\left. + \cos\left[\mathbf{G}_\nu \cdot \boldsymbol{r} + \mathbf{b}_\nu \cdot \left(\mathbf{s}^{(2)} - \mathbf{s}^{(3)}\right)\right] \right\}.$$

Note that even though the dependence on the stacking configurations varies on the atomic scale, the dependence on $\boldsymbol{r}$ is on the moiré scale. Finally, substituting the relations $\mathbf{s}^{(1)} = (-3\boldsymbol{u} + \boldsymbol{v} + 2\boldsymbol{w})/6$, $\mathbf{s}^{(2)} = (-\boldsymbol{v} + \boldsymbol{w})/3$, and $\mathbf{s}^{(3)} = (3\boldsymbol{u} + \boldsymbol{v} + 2\boldsymbol{w})/6$ in the above, we arrive at

$$\mathcal{V}_{ad}[\boldsymbol{r}, \boldsymbol{u}, \boldsymbol{v}] = V \sum_{\nu=1}^{3} \left\{ \cos\left[\frac{\mathbf{b}_\nu}{2} \cdot (\boldsymbol{v} - \boldsymbol{u}) - \mathbf{G}_\nu \cdot \boldsymbol{r}\right] \right.$$
$$\left. + \cos\left[\frac{\mathbf{b}_\nu}{2} \cdot (\boldsymbol{v} + \boldsymbol{u}) - \mathbf{G}_\nu \cdot \boldsymbol{r}\right] \right\}$$
$$= 2V \sum_{\nu=1}^{3} \cos\left(\frac{\mathbf{b}_\nu}{2}\boldsymbol{v} - \mathbf{G}_\nu\right) \cos\left(\frac{\mathbf{b}_\nu}{2}\boldsymbol{u}\right). \quad (S8)$$

We see that

$$\mathcal{V}_{ad}[\boldsymbol{r}, \boldsymbol{u}(\boldsymbol{r}), \boldsymbol{v}(\boldsymbol{r})] = \mathcal{V}_{ad}[\boldsymbol{r}, -\boldsymbol{u}(\boldsymbol{r}), \boldsymbol{v}(\boldsymbol{r})] \quad (S9)$$

as required by the mirror symmetry.

The net free energy is now given by the sum of Eqs. (S3) and (S4). By minimizing the free energy of variations with respect to $\boldsymbol{u}$ and $\boldsymbol{v}$, we obtain

$$\frac{\lambda + \mu}{2} \boldsymbol{\nabla}(\boldsymbol{\nabla} \cdot \boldsymbol{u}) + \frac{\mu}{2} \nabla^2 \boldsymbol{u} = \frac{\partial}{\partial \boldsymbol{u}} \mathcal{V}_{ad}[\boldsymbol{r}, \boldsymbol{u}(\boldsymbol{r}), \boldsymbol{v}(\boldsymbol{r})],$$

$$\frac{\lambda + \mu}{6} \boldsymbol{\nabla}(\boldsymbol{\nabla} \cdot \boldsymbol{v}) + \frac{\mu}{6} \nabla^2 \boldsymbol{v} = \frac{\partial}{\partial \boldsymbol{v}} \mathcal{V}_{ad}[\boldsymbol{r}, \boldsymbol{u}(\boldsymbol{r}), \boldsymbol{v}(\boldsymbol{r})]$$

$$\frac{\lambda + \mu}{4} \boldsymbol{\nabla}(\boldsymbol{\nabla} \cdot \boldsymbol{w}) + \frac{\mu}{4} \nabla^2 \boldsymbol{w} = 0. \quad (S10)$$

These equations describe a generalized two-dimensional version of the Frenkel-Kontorova model [9, 12] and will be important for understanding the effects of lattice relaxation later.

Our discussion so far has addressed the *static* problem. The *dynamical* equations of motion follow from the total Lagrangian $L = K - F$, where $F$ is defined in Eq. (S1) and the kinetic energy is given by

$$K = \frac{\rho}{2} \int d\boldsymbol{r} \sum_{l=1}^{3} \left(\dot{\mathbf{s}}^{(l)}\right)^2 = \frac{\rho}{2} \int d\boldsymbol{r} \left(\frac{\dot{\boldsymbol{u}}^2}{2} + \frac{\dot{\boldsymbol{v}}^2}{6} + \frac{\dot{\boldsymbol{w}}^2}{3}\right), \quad (S11)$$

with $\rho = 7.6 \times 10^{-7}$ kg/m² being the mass density of the individual graphene layers. Hereafter, we will not be concerned with the field $\boldsymbol{w}$, which just describes the in-phase displacements of the three layers, yielding the original acoustic phonons of graphene. Instead, focusing on the relative displacements, $\boldsymbol{u}$ and $\boldsymbol{v}$, we consider the deviations $\delta\boldsymbol{u}(\boldsymbol{r}, t) = \boldsymbol{u}(\boldsymbol{r}, t) - \boldsymbol{u}^{(0)}(\boldsymbol{r})$, $\delta\boldsymbol{v}(\boldsymbol{r}, t) = \boldsymbol{v}(\boldsymbol{r}, t) - \boldsymbol{v}^{(0)}(\boldsymbol{r})$, where $\boldsymbol{u}^{(0)}(\boldsymbol{r})$ and $\boldsymbol{v}^{(0)}(\boldsymbol{r})$ are metastable configurations satisfying Eq. (S10). Inserting these ansätze in Eqs. (S2) and (S4), we find that the free energy is of the form

$$F = F_{el}\left[\boldsymbol{u}^{(0)}, \boldsymbol{v}^{(0)}\right] + F_{el}\left[\delta\boldsymbol{u}, \delta\boldsymbol{v}\right] + F_{mix}\left[\delta\boldsymbol{u}, \boldsymbol{u}^{(0)}, \delta\boldsymbol{v}, \boldsymbol{v}^{(0)}\right] + F_{ad}\left[\delta\boldsymbol{u}, \boldsymbol{u}^{(0)}, \delta\boldsymbol{v}, \boldsymbol{v}^{(0)}\right], \quad (S12)$$

where the four terms are given by

$$F_{el}\left[\boldsymbol{u}^{(0)}, \boldsymbol{v}^{(0)}\right] = \int d\boldsymbol{r}\left[\frac{\lambda}{12}\left(3\partial_i u_i^{(0)} \partial_j u_j^{(0)} + \partial_i v_i^{(0)} \partial_j v_j^{(0)}\right) + \frac{\mu}{24}\left\{3\left(\partial_i u_j^{(0)} + \partial_j u_i^{(0)}\right)^2 + \left(\partial_i v_j^{(0)} + \partial_j v_i^{(0)}\right)^2\right\}\right],$$

$$F_{el}\left[\delta\boldsymbol{u}, \delta\boldsymbol{v}\right] = \int d\boldsymbol{r}\left[\frac{\lambda}{12}\left(3\partial_i \delta u_i \partial_j \delta u_j + \partial_i \delta v_i \partial_j \delta v_j\right) + \frac{\mu}{24}\left\{3\left(\partial_i \delta u_j + \partial_j \delta u_i\right)^2 + \left(\partial_i \delta v_j + \partial_j \delta v_i\right)^2\right\}\right],$$

$$F_{mix}\left[\delta\boldsymbol{u}, \boldsymbol{u}^{(0)}, \delta\boldsymbol{v}, \boldsymbol{v}^{(0)}\right] = \int d\boldsymbol{r}\left[\frac{\lambda}{2}\partial_i u_i^{(0)} \partial_j \delta u_j + \frac{\mu}{4}\left(\partial_i u_j^{(0)} + \partial_j u_i^{(0)}\right)\left(\partial_i \delta u_j + \partial_j \delta u_i\right)\right.$$
$$\left. + \frac{\lambda}{6}\partial_i v_i^{(0)} \partial_j \delta v_j + \frac{\mu}{12}\left(\partial_i v_j^{(0)} + \partial_j v_i^{(0)}\right)\left(\partial_i \delta v_j + \partial_j \delta v_i\right)\right], \quad (S13)$$

$$F_{\mathrm{ad}}\left[\delta\boldsymbol{u}, \boldsymbol{u}^{(0)}, \delta\boldsymbol{v}, \boldsymbol{v}^{(0)}\right] \simeq \int d\boldsymbol{r}\left[\mathcal{V}_{\mathrm{ad}}\left[\boldsymbol{r}, \boldsymbol{u}^{(0)}, \boldsymbol{v}^{(0)}\right] + \delta u_i \frac{\partial\mathcal{V}_{\mathrm{ad}}}{\partial u_i}\bigg|_{\boldsymbol{u}^{(0)}, \boldsymbol{v}^{(0)}} + \delta v_i \frac{\partial\mathcal{V}_{\mathrm{ad}}}{\partial v_i}\bigg|_{\boldsymbol{u}^{(0)}, \boldsymbol{v}^{(0)}}\right.$$
$$\left.+ \frac{1}{2}\delta u_i \delta u_j \frac{\partial^2\mathcal{V}_{\mathrm{ad}}}{\partial u_i \partial u_j}\bigg|_{\boldsymbol{u}^{(0)}, \boldsymbol{v}^{(0)}} + \frac{1}{2}\delta v_i \delta v_j \frac{\partial^2\mathcal{V}_{\mathrm{ad}}}{\partial v_i \partial v_j}\bigg|_{\boldsymbol{u}^{(0)}, \boldsymbol{v}^{(0)}} + \delta u_i \delta v_j \frac{\partial^2\mathcal{V}_{\mathrm{ad}}}{\partial u_i \partial v_j}\bigg|_{\boldsymbol{u}^{(0)}, \boldsymbol{v}^{(0)}}\right]. \tag{S14}$$

In particular, $F_{\mathrm{ad}}$ in Eq. (S14) results from expanding the adhesion energy as a Taylor series up to quadratic order.

A reorganization of the terms above now proves convenient. Upon integrating by parts, we notice that

$$F \simeq F_0 + U\left[\delta\boldsymbol{u}, \delta\boldsymbol{v}\right] + \int d\boldsymbol{r}\,\delta u_i\left[\frac{\partial\mathcal{V}_{\mathrm{ad}}}{\partial u_i}\bigg|_{\boldsymbol{u}^{(0)}, \boldsymbol{v}^{(0)}} - \frac{\lambda}{2}\partial_i\partial_j u_j^{(0)} - \frac{\mu}{2}\partial_j\left(\partial_i u_j^{(0)} + \partial_j u_i^{(0)}\right)\right]$$
$$+ \int d\boldsymbol{r}\,\delta v_i\left[\frac{\partial\mathcal{V}_{\mathrm{ad}}}{\partial v_i}\bigg|_{\boldsymbol{u}^{(0)}, \boldsymbol{v}^{(0)}} - \frac{\lambda}{6}\partial_i\partial_j v_j^{(0)} - \frac{\mu}{6}\partial_j\left(\partial_i v_j^{(0)} + \partial_j v_i^{(0)}\right)\right]. \tag{S15}$$

The first term in Eq. (S15), $F_0$, is simply the free energy of the equilibrium solution,

$$F_0 = \int d\boldsymbol{r}\left\{\frac{\lambda}{4}\left(\boldsymbol{\nabla}\cdot\boldsymbol{u}^{(0)}\right)^2 + \frac{\mu}{8}\left(\partial_i u_j^{(0)} + \partial_j u_i^{(0)}\right)^2 + \frac{\lambda}{12}\left(\boldsymbol{\nabla}\cdot\boldsymbol{v}^{(0)}\right)^2 + \frac{\mu}{24}\left(\partial_i v_j^{(0)} + \partial_j v_i^{(0)}\right)^2 + \mathcal{V}_{\mathrm{ad}}\left[\boldsymbol{r}, \boldsymbol{u}^{(0)}\left(\boldsymbol{r}\right), \boldsymbol{v}^{(0)}\left(\boldsymbol{r}\right)\right]\right\},$$

whereas the second, $U[\delta\boldsymbol{u}, \delta\boldsymbol{v}]$, describes the spectrum of harmonic oscillations,

$$U\left[\delta\boldsymbol{u}, \delta\boldsymbol{v}\right] = \int d\boldsymbol{r}\left[\frac{\lambda}{4}\left(\boldsymbol{\nabla}\cdot\delta\boldsymbol{u}\right)^2 + \frac{\mu}{8}\left(\partial_i\delta u_j + \partial_j\delta u_i\right)^2 + \frac{\lambda}{12}\left(\boldsymbol{\nabla}\cdot\delta\boldsymbol{v}\right)^2 + \frac{\mu}{24}\left(\partial_i\delta v_j + \partial_j\delta v_i\right)^2\right.$$
$$\left.+ \frac{1}{2}\delta u_i \delta u_j \frac{\partial^2\mathcal{V}_{\mathrm{ad}}}{\partial u_i \partial u_j}\bigg|_{\boldsymbol{u}^{(0)}, \boldsymbol{v}^{(0)}} + \frac{1}{2}\delta v_i \delta v_j \frac{\partial^2\mathcal{V}_{\mathrm{ad}}}{\partial v_i \partial v_j}\bigg|_{\boldsymbol{u}^{(0)}, \boldsymbol{v}^{(0)}} + \delta u_i \delta v_j \frac{\partial^2\mathcal{V}_{\mathrm{ad}}}{\partial u_i \partial v_j}\bigg|_{\boldsymbol{u}^{(0)}, \boldsymbol{v}^{(0)}}\right]. \tag{S16}$$

Finally, the last two terms on the right in Eq. (S15) are identically zero since the integrands vanish by virtue of Eq. (S10). Then, from Eqs. (S11) and (S16), we obtain the coupled Euler-Lagrange equations

$$-\frac{\rho}{2}\delta\ddot{u}_i + \frac{\lambda+\mu}{2}\partial_i\partial_j\delta u_j + \frac{\mu}{2}\partial_j\partial_j\delta u_i = \delta u_j \frac{\partial^2\mathcal{V}_{\mathrm{ad}}}{\partial u_i \partial u_j}\bigg|_{\boldsymbol{u}^{(0)}, \boldsymbol{v}^{(0)}} + \delta v_j \frac{\partial^2\mathcal{V}_{\mathrm{ad}}}{\partial u_i \partial v_j}\bigg|_{\boldsymbol{u}^{(0)}, \boldsymbol{v}^{(0)}}, \tag{S17}$$

$$-\frac{\rho}{6}\delta\ddot{v}_i + \frac{\lambda+\mu}{6}\partial_i\partial_j\delta v_j + \frac{\mu}{6}\partial_j\partial_j\delta v_i = \delta v_j \frac{\partial^2\mathcal{V}_{\mathrm{ad}}}{\partial v_i \partial v_j}\bigg|_{\boldsymbol{u}^{(0)}, \boldsymbol{v}^{(0)}} + \delta u_j \frac{\partial^2\mathcal{V}_{\mathrm{ad}}}{\partial v_i \partial u_j}\bigg|_{\boldsymbol{u}^{(0)}, \boldsymbol{v}^{(0)}}. \tag{S18}$$

These Euler-Lagrange equations, which constitute the central result of this section, will be used extensively to obtain the phason spectra in the rest of this work. Due to the constraint of mirror symmetry in Eq. (S9), we have

$$\frac{\partial^2\mathcal{V}_{\mathrm{ad}}}{\partial v_i \partial u_j}\bigg|_{\boldsymbol{u}^{(0)}, \boldsymbol{v}^{(0)}} = -\frac{\partial^2\mathcal{V}_{\mathrm{ad}}}{\partial v_i \partial u_j}\bigg|_{-\boldsymbol{u}^{(0)}, \boldsymbol{v}^{(0)}}. \tag{S19}$$

### A. Solutions of the Euler-Lagrange equations

The solutions to Eqs. (S17) and (S18) are more easily obtained in Fourier space. We introduce Fourier series for the field $\boldsymbol{u}$ (and analogously for $\boldsymbol{v}$) using the conventions,

$$\boldsymbol{u}^{(0)}(\boldsymbol{r}) = \sum_{\mathbf{G}}\boldsymbol{u}_{\mathbf{G}}^{(0)}\,e^{i\mathbf{G}\cdot\boldsymbol{r}}, \quad \delta\boldsymbol{u}(\boldsymbol{r}, t) = \int\frac{d\omega}{2\pi}\sum_{\boldsymbol{q}}\delta\boldsymbol{u}_{\boldsymbol{q}}(\omega)\,e^{i\boldsymbol{q}\cdot\boldsymbol{r} - i\omega t}, \tag{S20}$$

where $\boldsymbol{q}$ lies in the monolayer Brillouin zone. Here, we have implicitly assumed that $\boldsymbol{u}^{(0)}$ varies on the moiré scale, so it admits a Fourier expansion in $\{\boldsymbol{G}\}$ which are the reciprocal lattice vectors of the moiré lattice. We also define the Fourier components for derivatives of the adhesion potential,

$$\frac{\partial}{\partial u_i}\mathcal{V}_{\mathrm{ad}}\bigg|_{\boldsymbol{u}^{(0)}, \boldsymbol{v}^{(0)}} \equiv \sum_{\nu, \mathbf{G}}\mathcal{F}_{\mathbf{G}}^{\nu}\,e^{i\mathbf{G}\cdot\boldsymbol{r}}\,b_{\nu, i}, \qquad \frac{\partial}{\partial v_i}\mathcal{V}_{\mathrm{ad}}\bigg|_{\boldsymbol{u}^{(0)}, \boldsymbol{v}^{(0)}} \equiv \sum_{\nu, \mathbf{G}}\tilde{\mathcal{F}}_{\mathbf{G}}^{\nu}\,e^{i\mathbf{G}\cdot\boldsymbol{r}}\,b_{\nu, i}, \tag{S21}$$

$$\frac{\partial^2 \mathcal{V}_{\rm ad}}{\partial u_i \partial u_j}\Big|_{\boldsymbol{u}^{(0)}, \boldsymbol{v}^{(0)}} \equiv \sum_{\nu, \mathbf{G}} \mathcal{J}_{\mathbf{G}}^\nu \, e^{i\mathbf{G}\cdot\boldsymbol{r}} \, b_{\nu, i} \, b_{\nu, j}, \quad \frac{\partial^2 \mathcal{V}_{\rm ad}}{\partial v_i \partial v_j}\Big|_{\boldsymbol{u}^{(0)}, \boldsymbol{v}^{(0)}} \equiv \sum_{\nu, \mathbf{G}} \tilde{\mathcal{J}}_{\mathbf{G}}^\nu \, e^{i\mathbf{G}\cdot\boldsymbol{r}} \, b_{\nu, i} \, b_{\nu, j}, \quad \frac{\partial^2 \mathcal{V}_{\rm ad}}{\partial u_i \partial v_j}\Big|_{\boldsymbol{u}^{(0)}, \boldsymbol{v}^{(0)}} \equiv \sum_{\nu, \mathbf{G}} \mathcal{K}_{\mathbf{G}}^\nu \, e^{i\mathbf{G}\cdot\boldsymbol{r}} \, b_{\nu, i} \, b_{\nu, j}.$$
(S22)

With these definitions, the Euler-Lagrange equations can be solved as follows. First, from Eq. (S10), one can obtain the static configurations $\boldsymbol{u}_{\mathbf{G}}^{(0)}, \boldsymbol{v}_{\mathbf{G}}^{(0)}$ as

$$\boldsymbol{u}_{\mathbf{G}}^{(0)} = -2\sum_{\nu=1}^{3} \mathcal{F}_{\mathbf{G}}^\nu \, \hat{H}_{\mathbf{G}}^{-1} \, \mathbf{b}_\nu, \quad \boldsymbol{v}_{\mathbf{G}}^{(0)} = -6\sum_{\nu=1}^{3} \tilde{\mathcal{F}}_{\mathbf{G}}^\nu \, \hat{H}_{\mathbf{G}}^{-1} \, \mathbf{b}_\nu \quad \text{with } \hat{H}_{\mathbf{G}} = \begin{bmatrix} (\lambda + 2\mu)\, G_x^2 + \mu G_y^2 & (\lambda + \mu)\, G_x G_y \\ (\lambda + \mu)\, G_x G_y & (\lambda + 2\mu)\, G_y^2 + \mu G_x^2 \end{bmatrix}.$$
(S23)

As we show below, in the absence of external mirror-symmetry breaking, we find $\boldsymbol{u}^{(0)} = 0$, indicating that the mirror-symmetric configuration is energetically most favorable, in accordance with experiments [11]. Consequently, we see that the last terms in both Eqs. (S17) and (S18) vanish, implying that the modes described by $\delta \boldsymbol{u}$ and $\delta \boldsymbol{v}$ do not mix within the quadratic approximation, as expected since they are odd and even under the mirror symmetry, respectively.

Using the thus-obtained relaxation textures to compute $\mathcal{J}_{\mathbf{G}}^\nu, \tilde{\mathcal{J}}_{\mathbf{G}}^\nu, \mathcal{K}_{\mathbf{G}}^\nu$, the problem reduces to solving the following secular equations

$$\rho\, \omega^2 \, \delta\boldsymbol{u}_{\mathbf{G}+\boldsymbol{q}}(\omega) = \hat{H}_{\mathbf{G}+\boldsymbol{q}}\, \delta\boldsymbol{u}_{\mathbf{G}+\boldsymbol{q}}(\omega) + 2\sum_{\mathbf{G}'}\sum_{\nu=1}^{3} \left( \mathcal{J}_{\mathbf{G}-\mathbf{G}'}^\nu \, \hat{B}_\nu \, \delta\boldsymbol{u}_{\mathbf{G}'+\boldsymbol{q}}(\omega) + \mathcal{K}_{\mathbf{G}-\mathbf{G}'}^\nu \, \hat{B}_\nu \, \delta\boldsymbol{v}_{\mathbf{G}'+\boldsymbol{q}}(\omega) \right),$$
(S24a)

$$\rho\, \omega^2 \, \delta\boldsymbol{v}_{\mathbf{G}+\boldsymbol{q}}(\omega) = \hat{H}_{\mathbf{G}+\boldsymbol{q}}\, \delta\boldsymbol{v}_{\mathbf{G}+\boldsymbol{q}}(\omega) + 6\sum_{\mathbf{G}'}\sum_{\nu=1}^{3} \left( \tilde{\mathcal{J}}_{\mathbf{G}-\mathbf{G}'}^\nu \, \hat{B}_\nu \, \delta\boldsymbol{v}_{\mathbf{G}'+\boldsymbol{q}}(\omega) + \mathcal{K}_{\mathbf{G}-\mathbf{G}'}^\nu \, \hat{B}_\nu \, \delta\boldsymbol{u}_{\mathbf{G}'+\boldsymbol{q}}(\omega) \right),$$
(S24b)

where we have defined the matrix

$$\hat{B}_\nu = \begin{bmatrix} b_{\nu, x} b_{\nu, x} & b_{\nu, x} b_{\nu, y} \\ b_{\nu, x} b_{\nu, y} & b_{\nu, y} b_{\nu, y} \end{bmatrix}.$$
(S25)

Note that in the mirror-symmetric limit, $\mathcal{K} = 0$ due to Eq. (S19); however, we retain these terms in Eq. (S24) as they will be nonzero when we study cases with broken mirror symmetry below.

### B. Relation between phonon modes of TBG and TTG

We now show that the lattice relaxation in TTG and TBG, as well as their phason modes, can be related by a rescaling of parameters. To this end, let us define $\boldsymbol{x}^{(0)} = \boldsymbol{v}^{(0)}/2$ which—upon recalling that $\boldsymbol{u}^{(0)} = 0$ when mirror symmetry is preserved—allows us to rewrite the second line in Eq. (S10) as

$$\frac{\lambda + \mu}{3}\, \boldsymbol{\nabla}\left( \boldsymbol{\nabla} \cdot \boldsymbol{x}^{(0)} \right) + \frac{\mu}{3}\, \nabla^2 \boldsymbol{x}^{(0)} = \frac{\partial}{\partial \boldsymbol{x}} V \sum_\nu \cos\left( \mathbf{b}_\nu \boldsymbol{x} - \mathbf{G}_\nu \boldsymbol{r} \right)\Big|_{\boldsymbol{x}=\boldsymbol{x}^{(0)}}.$$
(S26)

Similarly, defining $\delta\boldsymbol{x} \equiv \boldsymbol{v}/2 - \boldsymbol{x}^{(0)}$, we can write the Euler-Lagrange equation (S18) as

$$-\rho\, \delta\ddot{x}_i + (\lambda + \mu)\partial_i \partial_j \delta x_j + \mu\, \partial_j \partial_j \delta x_i = \delta x_j \frac{\partial^2}{\partial x_i \partial x_j} 3V \sum_\nu \cos\left( \mathbf{b}_\nu \boldsymbol{x} - \mathbf{G}_\nu \boldsymbol{r} \right)\Big|_{\boldsymbol{x}=\boldsymbol{x}^{(0)}}.$$
(S27)

Comparison of the corresponding equations for the phason mode of TBG [2] reveals that they are identical upon rescaling $V \to V/3$, $\rho \to \rho/2$, $\lambda \to \lambda/2$, and $\mu \to \mu/2$ in Eqs. (S26) and (S27). In other words, denoting the lattice relaxation of TBG for some given $V$, $\rho$, $\lambda$, and $\mu$ by $\boldsymbol{v}_{\rm TBG}^{(0)}(\boldsymbol{r}; V, \rho, \lambda, \mu)$ and the phason dispersion by $\omega_{\rm TBG}(\boldsymbol{q}; V, \rho, \lambda, \mu)$, it holds that the corresponding $\boldsymbol{v}^{(0)}$

and $\omega_{\rm phason}$ for TTG are simply

$$\boldsymbol{v}^{(0)}(\boldsymbol{r}; V, \rho, \lambda, \mu) = 2\boldsymbol{v}_{\rm TBG}^{(0)}(\boldsymbol{r}; 3V, 2\rho, 2\lambda, 2\mu),$$
(S28)

$$\omega_{\rm phason}(\boldsymbol{q}; V, \rho, \lambda, \mu) = \omega_{\rm TBG}(\boldsymbol{q}; 3V, 2\rho, 2\lambda, 2\mu).$$
(S29)

Naturally, neither $\boldsymbol{v}^{(0)}$ nor $\omega_{\rm phason}$ change when rescaling all parameters $V$, $\rho$, $\lambda$, and $\mu$ such that there are several

ways of expressing the rescaling from TTG to TBG. A particularly compact one is $V \to 2V/3$, as stated in the main text.

Apart from explaining the similarities of the phason spectrum shown for TTG in the last panel of Fig. 2(a) to that of TBG [1, 2], these relations tell us about the angle dependence of the strength of lattice relaxation. A natural length scale of the stacking textures in TBG is given by [2]

$$\ell_{\text{TBG}} = \frac{a}{\pi} \sqrt{\frac{\mu}{2V}}, \qquad (S30)$$

which captures the energetic compromise between the adhension potential and intralayer elastic energy cost; we now immediately see from the above-mentioned rescaling that

$$\ell_{\text{TTG}} = \sqrt{\frac{2}{3}} \ell_{\text{TBG}}. \qquad (S31)$$

Since lattice relaxation becomes sizable when this length scale is comparable to or smaller than the moiré length $L_M$, we conclude that lattice relaxation at a given twist angle is stronger in TTG than in TBG, which agrees with first-principle calculations [13]. More explicitly, since $L_M \propto 1/\theta$, we see that the characteristic angles $\theta_{\text{TBG}}^*$ and $\theta_{\text{TTG}}^*$ below which lattice relaxation becomes important for TBG and TTG, respectively, are related as $\theta_{\text{TTG}}^* \propto \sqrt{3/2}\, \theta_{\text{TBG}}^*$. Note that the magic angle is larger by a factor of about $\sqrt{2} > \sqrt{3/2}$ in TTG, so lattice relaxation is slightly weaker in TTG than in TBG at their respective magic angles.

### C. Flexural phonons

In our analysis so far, we have focused on *in-plane* displacements of the graphene layers and, thus, neglected flexural phonon modes. By generalizing the arguments of Ref. [1] to the trilayer case, we show here that these modes are not affected by the moiré-superlattice modulation and, therefore, are largely independent of twist angle.

To capture out-of-plane modes, we have to introduce additional out-of-plane displacements, $s_z^{(j)}$, for each of the three layers $l = 1, 2, 3$. The interlayer binding potential can be approximated by [1]

$$F_\perp = \int d\boldsymbol{r}\, \mathcal{V}_\perp[s_z^{(l)}(\boldsymbol{r}), \boldsymbol{\Delta}_j(\boldsymbol{r})],$$

$$\mathcal{V}_\perp[s_z^{(l)}, \boldsymbol{\Delta}_j] = V_\perp \left( h_0 + s_z^{(2)} - s_z^{(1)} - h_z(\boldsymbol{\Delta}_1) \right)^2 \qquad (S32)$$
$$+ V_\perp \left( h_0 + s_z^{(3)} - s_z^{(2)} - h_z(\boldsymbol{\Delta}_3) \right)^2,$$

where $V_\perp > 0$, and $h_z(\boldsymbol{\Delta})$ denotes the energetically most favorable distance between the graphene layers for a local stacking parameterized by sliding vector $\boldsymbol{\Delta}$, and $h_0 =$

$h_z(0)$ is a reference height. Note that mirror symmetry acts as $s_z^{(1,3)} \to -s_z^{(3,1)}$, $s_z^{(2)} \to -s_z^{(2)}$, and $\boldsymbol{\Delta}_{1,3} \to \boldsymbol{\Delta}_{3,1}$; as required, $F_\perp$ is invariant under it. To proceed, we reparameterize $\{s_z^{(l)}\} \to \{d_1, d_2, w_z\}$ with $d_1 = s_z^{(2)} - s_z^{(1)}$, $d_3 = s_z^{(3)} - s_z^{(2)}$, and $w_z = s_z^{(1)} + s_z^{(2)} + s_z^{(3)}$, leading to

$$\mathcal{V}_\perp = V_\perp \sum_{j=1,3} \left[ h_0 + d_j(\boldsymbol{r}) - h_z(\boldsymbol{\Delta}_j(\boldsymbol{r})) \right]^2. \qquad (S33)$$

First, we see that, similar to the in-plane acoustic modes—where all three graphene layers move together, and which are thus unaffected by $F_{\text{ad}}$—the out-of-plane modes associated with $w_z$ do not "feel" $F_\perp$; as such, they are independent of the twist angle. To study the phonon modes associated with $d_j$, we write $d_j(\boldsymbol{r}, t) = d_j^{(0)}(\boldsymbol{r}) + \delta d_j(\boldsymbol{r}, t)$, where $\delta d_j$ are the fluctuations away from the equilibrium configuration $d_j^{(0)}(\boldsymbol{r})$. As follows from Eq. (S33), the latter is approximately given by $d_j^{(0)}(\boldsymbol{r}) \simeq h_0 - h_z(\boldsymbol{\Delta}_j^{(0)}(\boldsymbol{r}))$ [1], where $\boldsymbol{\Delta}_j^{(0)}$ is $\boldsymbol{\Delta}_j$ evaluated at $\boldsymbol{u} = \boldsymbol{u}^{(0)}$, $\boldsymbol{v} = \boldsymbol{v}^{(0)}$. Neglecting the difference between $\boldsymbol{\Delta}_j^{(0)}$ and $\boldsymbol{\Delta}_j$, $\mathcal{V}_\perp \simeq V_\perp \sum_{j=1,3} (\delta d_j)^2$ and the moiré superlattice does not affect these phonon modes either. Taking into account that $\boldsymbol{\Delta}_j \neq \boldsymbol{\Delta}_j^{(0)}$ in the presence of in-plane phonons will lead to a coupling of the in-plane modes and $\delta d_j$. We neglect this coupling as the $\delta d_j$ modes have a finite gap (of order $100\,\text{cm}^{-1}$ [14]) which is larger than the low-energy range ($< 50\,\text{cm}^{-1}$) that we are interested in; therefore, this coupling is expected to be negligible for our results.

## SII. PHONON SPECTRA

In the course of this section, we employ the formalism developed in Sec. SI to compute the lattice relaxation and phonon spectra in TTG, both with and without external mirror-symmetry breaking, for a wide range of twist angles. However, we do not yet consider the effects of broken electronic symmetries, which will be addressed in the next section.

### A. Importance of lattice relaxation

As discussed in the main text, the textures of $\boldsymbol{u}^{(0)}(\boldsymbol{r})$ and $\boldsymbol{v}^{(0)}(\boldsymbol{r})$ together encode the relaxation of the moiré superlattice. To pick a starting point, the simplest possible approximation that one can make is to neglect all lattice relaxation effects and set $\boldsymbol{u}^{(0)} = \boldsymbol{v}^{(0)} = 0 \; \forall \; \boldsymbol{r}$. In such a scenario—which describes a case where the three graphene layers are "floating" on top of one another—one may naively imagine that the mirror-even mode obtained by displacing the middle layer against the outer two simply corresponds to a translation of the moiré superlattice and is thus gapless. However, we find that this simplified model incorrectly predicts a strong (and

unphysical) softening of the acoustic branches near the Brillouin zone center as demonstrated by Fig. S1, which illustrates the phonon spectrum of unrelaxed TTG for both large ($\theta = 3\,\theta_m$) and small ($\theta = \theta_m$) twist angles, with $\theta_m = 1.56°$ being the magic angle. These unstable phonon modes, with large imaginary frequencies at the $\Gamma$ point in the dispersion, convey a structural instability of the rigidly twisted trilayer, highlighting that the relaxation of the twisted structure is necessary to obtain a suitable ground state about which one can consider vibrations.

Subsequently incorporating lattice relaxation effects in computing the phonon frequencies, we find that the spatial reorganization of atoms within the graphene sheets stabilizes the heterostructure, removing the unstable phonon modes in the process (see Fig. S3). In particular, we observe that this atomic reconstruction leads to a nontrivial $\boldsymbol{v}^{(0)} \neq 0$ and the stacking texture becomes progressively sharper for smaller twist angles [Fig. S2(a-d)]. At the same time, $\boldsymbol{u}^{(0)}(\boldsymbol{r})$ is seen to be identically zero $\forall\ \boldsymbol{r}$: the form in the last line of Eq. (S8) provides a more direct understanding of this finding that there is no spontaneous lattice relaxation in the mirror-odd channel—in agreement with experiment [11] and previous relaxation calculations [15]. To see that Eq. (S8) implies that $\boldsymbol{u}^{(0)} = 0$ is at least a local minimum, let us first set $\boldsymbol{u} = 0$. The energetics determining $\boldsymbol{v}^{(0)}$ are now of the same form as in TBG (as argued in Sec. SIB), so we expect a texture for $\boldsymbol{v}^{(0)}$ that optimizes the trade-off between minimizing the intralayer elastic energy and maximizing the regions of AB/BA stacking between the outer and the inner layer. As $\boldsymbol{v}^{(0)} = 0$ will lead to a vanishing spatial average

$$\alpha_\nu = \frac{1}{\Omega} \int \mathrm{d}^2\boldsymbol{r}\,\cos\left(\mathbf{b}_\nu \boldsymbol{v}^{(0)}(\boldsymbol{r})/2 - \mathbf{G}_\nu \boldsymbol{r}\right) \quad \text{(S34)}$$

over the system (with area $\Omega$) without any intralayer energetic cost, we must have $\alpha_\nu < 0$ for $\boldsymbol{v}^{(0)} \neq 0$ minimizing the elastic energy. Bearing this in mind, we see, from Eq. (S8), that turning on a finite value of $\boldsymbol{u}^{(0)}$ would immediately decrease the energetic gain from the mirror-even relaxation and, at the same time, cost intralayer elastic energy [see Eq. (S3)], which is why $\boldsymbol{u}^{(0)} = 0$ is favored. More intuitively, the mirror-symmetric configuration is naturally preferred as it allows the system to maximize the AB/BA stacking regions of the top-middle and middle-bottom pair simultaneously by a mirror-even relaxation.

## B. Gap of the mirror-odd shear mode

Figure S3 arrays the spectra of the phasons and the gapped shear modes of TTG once lattice relaxation effects have been taken into account. The velocities and gaps of these two modes, respectively, as a function of twist angle are sketched in Figs. 2(c-e). While the phason mode is gapless, to gain insight into the behavior of the gap of the mirror-odd shear modes $\delta\boldsymbol{u}$, here, we apply perturbation theory. To this end, we generalize the form of the adhesion potential in Eq. (S8) to the family

$$V \sum_{\nu,p=\pm} \cos\left[\frac{\mathbf{b}_\nu}{2}\left(\boldsymbol{v} + \varepsilon\, p\, \boldsymbol{u}\right) - \mathbf{G}_\nu \boldsymbol{r}\right] \quad \text{(S35)}$$

and treat $\varepsilon > 0$ as a small parameter (where $\varepsilon = 1$ corresponds to the actual system). Importantly, this generalization does not break any physical symmetry. Further, note that we still have $\boldsymbol{u}^{(0)} = 0$ for any $\varepsilon \leq 1$ (since $\varepsilon < 1$ will not make it more favorable energetically to break mirror symmetry); consequently, $\boldsymbol{v}^{(0)}$ also does not depend on $\varepsilon$ (and neither does any other property of the phason sector).

Using the multi-index notation $(\delta \boldsymbol{\mathcal{U}}_{\boldsymbol{q}})_{j,\mathbf{G}} = \delta(\boldsymbol{u}_{\mathbf{G}+\boldsymbol{q}})_j$ and recalling that $\mathcal{K} = 0$ in the mirror-symmetric limit, we can rewrite Eq. (S24a) as $\mathcal{H}_{\boldsymbol{q}}\delta\boldsymbol{\mathcal{U}}_{\boldsymbol{q}} = \rho\,\omega_{\boldsymbol{q}}^2\delta\boldsymbol{\mathcal{U}}_{\boldsymbol{q}}$, where

$$(\mathcal{H}_{\boldsymbol{q}})_{j,\mathbf{G};j',\mathbf{G}'} = \delta_{\mathbf{G},\mathbf{G}'}(\hat{H}_{\mathbf{G}+\boldsymbol{q}})_{j,j'} + 2\varepsilon^2\sum_{\nu=1}^{3}\mathcal{J}_{\mathbf{G}-\mathbf{G}'}^\nu(B_\nu)_{j,j'}.$$

On treating $\varepsilon^2$ as a perturbation, the zeroth-order eigenstates are of the form $(\delta\boldsymbol{\mathcal{U}}_{\boldsymbol{q}}^{(0)})_{j,\mathbf{G}} = (e_{\mathbf{G}+\boldsymbol{q}}^\pm)_j\delta_{\mathbf{G},\mathbf{G}_0}$ with

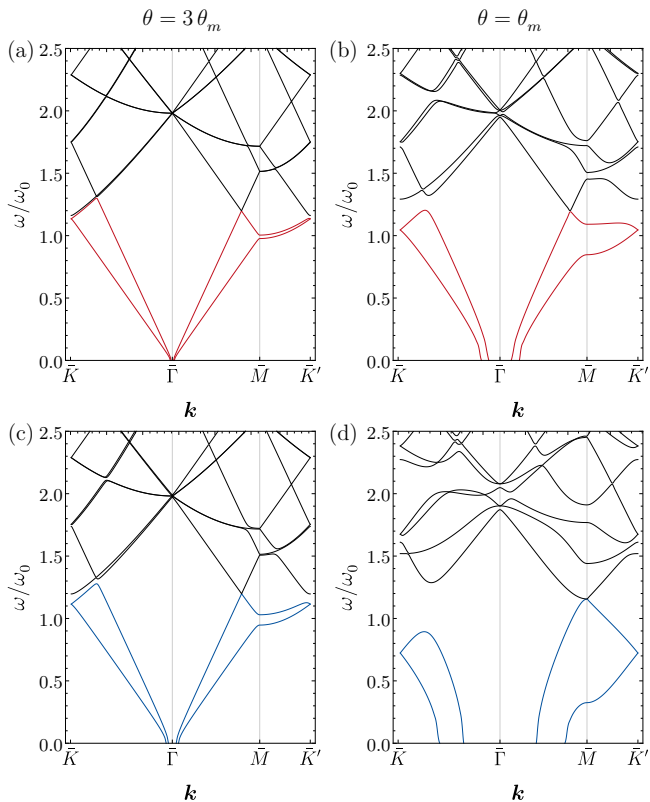

FIG. S1. Spectra in the (a,b) mirror-odd and (c,d) mirror-even sectors for two different twist angles when neglecting lattice relaxation. Imaginary frequencies indicate structural instabilities, which are more prominent at smaller twist angles.

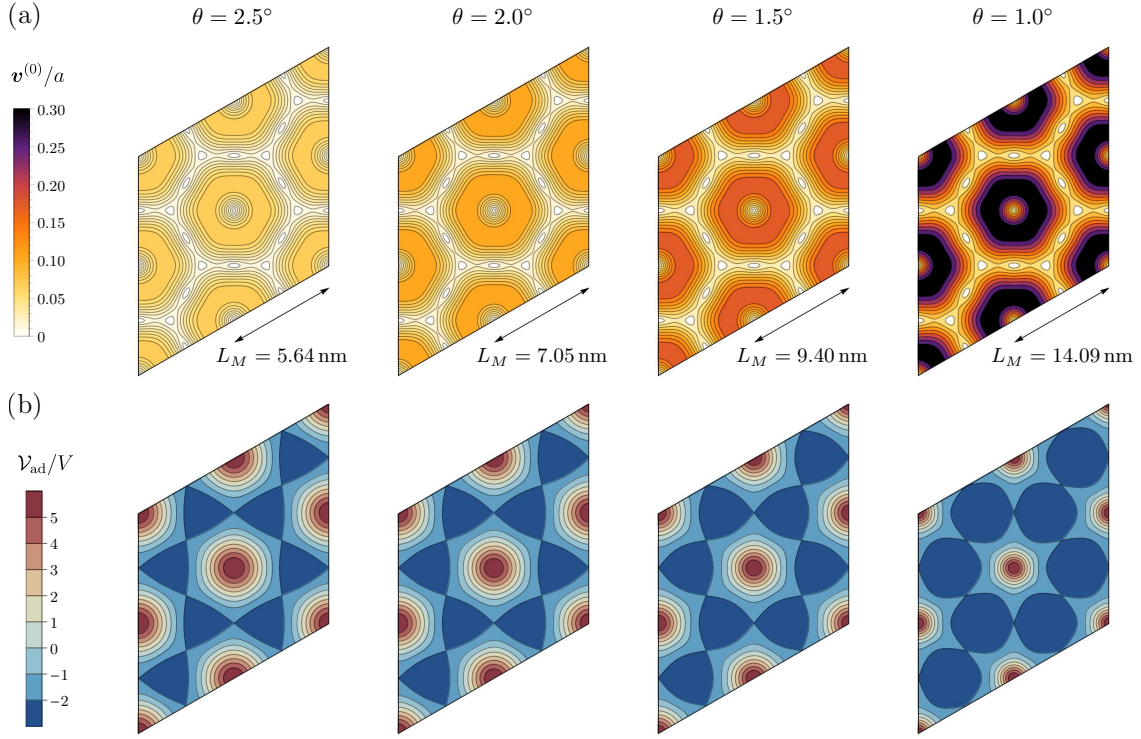

FIG. S2. Spatial dependence of the relaxation for different twist angles. In the presence of mirror symmetry, we have only relaxation in the mirror-even sector. The regions of AB/BA stacking (dark blue) expand whereas the regions of AA stacking (at the center of the rhombi) shrink.

eigenvalues $(3\mu+\lambda\pm(\mu+\lambda))(\mathbf{G}_0+\boldsymbol{q})^2/2$. We see that, for $\boldsymbol{q}$ in the moiré Brillouin zone, the lowest-energy solutions correspond to $\mathbf{G}_0 = 0$.

To compute the gap, we focus on $\boldsymbol{q} = 0$, where the two unperturbed lowest-energy solutions, with polarizations $\boldsymbol{e}_{\mathbf{G}_0=0}^{\pm}$, are degenerate but separated by a finite gap to the other states (with $\mathbf{G}_0 \neq 0$). Therefore, to first order in $V$, the energy of the mirror-odd shear modes are given by the eigenvalues of $\mathcal{H}_{\boldsymbol{q}=0}$ projected to this two dimensional subspace,

$$\left(\mathcal{H}^{\text{eff}}\right)_{j,j'} = 2\varepsilon^2 \sum_{\nu=1}^{3} \mathcal{J}_{\mathbf{G}=0}^{\nu}(B_\nu)_{j,j'} = -\varepsilon^2 \delta_{j,j'} \frac{3V}{2} |\mathbf{b}|^2 \alpha;$$

in this equation, $|\mathbf{b}| \equiv |\mathbf{b}_\nu|$ and $\alpha \equiv \alpha_\nu \leq 0$, with $\alpha_\nu$ as defined in Eq. (S34), which does not depend on $\nu$ due to rotational symmetry.

Therefore, the two modes remain degenerate but acquire a finite energy gap given by Eq. (3) of the main text. The fact that the effective Hamiltonian $\mathcal{H}^{\text{eff}}$ is diagonal is simply a consequence of $C_3$ rotational symmetry; as such, the degeneracy of the lowest-energy mirror-odd shear modes at $\boldsymbol{q}=0$—transforming under the $E$ representation of $C_3$—holds to arbitrary order in $V$ as long as $C_3$ and the gap to the higher-energy modes remain intact.

## C. Effect of a displacement field

The perpendicular displacement field, $D_0$, in Fig. 1 is odd under mirror reflections $\sigma_h$, so a finite $D_0 \neq 0$ breaks $\sigma_h$. As a result, in the presence of such an applied field, the Lamé coefficients of the elastic theory in Eq. (S2) will, in general, be of the form

$$\begin{aligned}
\lambda_1 &= \lambda_a - c_\lambda D_0 + \mathcal{O}(D_0^2), \\
\lambda_2 &= \lambda_b + \mathcal{O}(D_0^2), \\
\lambda_3 &= \lambda_a + c_\lambda D_0 + \mathcal{O}(D_0^2),
\end{aligned} \tag{S36}$$

and likewise for $\mu_l$. Furthermore, the parameters of the adhesion potential (S7) are also allowed to be nonuniform, according as

$$\begin{aligned}
V_1 &= V - \gamma D_0 + \mathcal{O}(D_0^2), \\
V_3 &= V + \gamma D_0 + \mathcal{O}(D_0^2).
\end{aligned} \tag{S37}$$

The incorporation of the varying Lamé coefficients (S36) lies beyond the scope of our previously derived Euler-Lagrange equations (S24). Thus, in our framework, we theoretically model the effect of the displacement field by Eq. (S37), treating $\gamma$ as a phenomenological parameter. Since $V_1 \neq V_3$ leads to mirror-odd lattice relaxation, $\boldsymbol{u}^{(0)} \neq 0$, the coefficients $\mathcal{K}_{\mathbf{G}}^{\nu}$ in Eq. (S24) are now nonzero, so the equations of motion for $\delta\boldsymbol{u}$

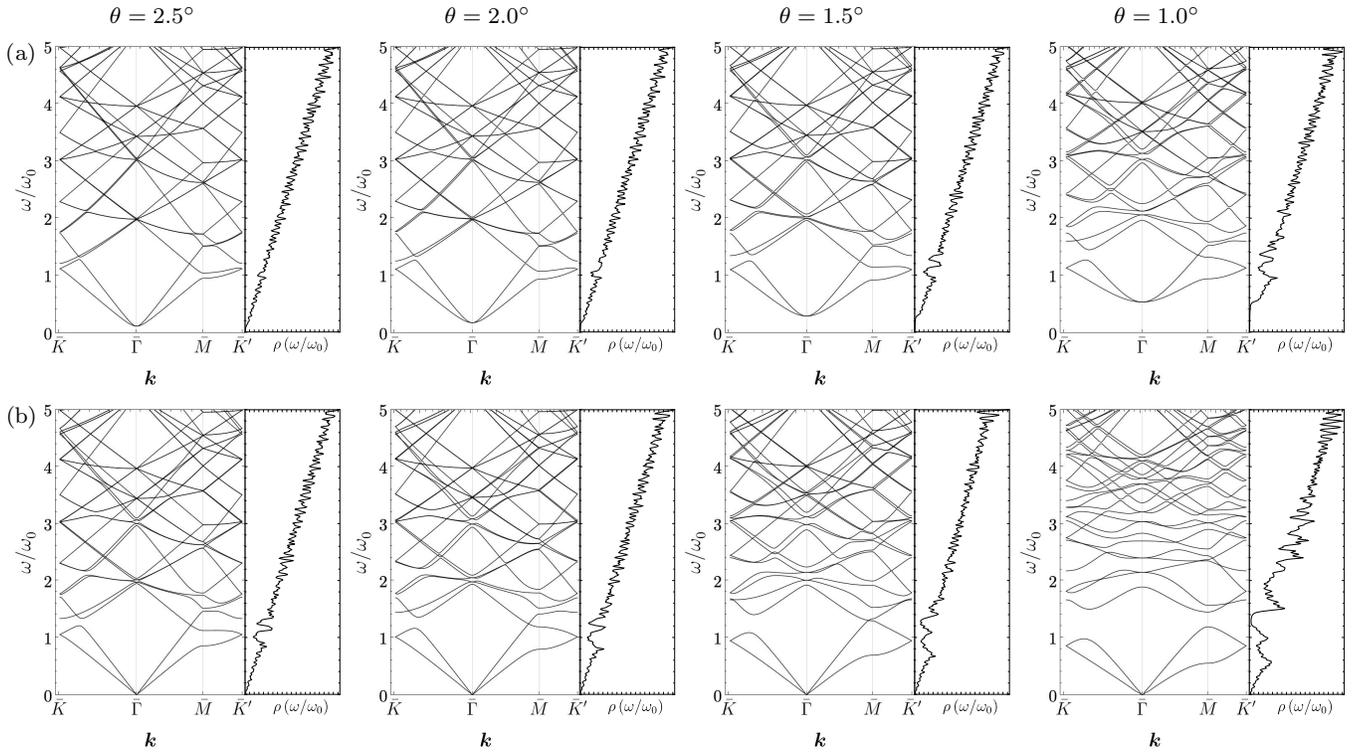

FIG. S3. (a) Phonon dispersions for the mirror-odd gapped shear mode $\boldsymbol{u}$ as a function of the twist angle $\theta$. The horizontal axis traces a path connecting high-symmetry points in the moiré Brillouin zone (MBZ). The right panels plot the density of states $\rho$ as a function of the dimensionless frequency $\omega/\omega_0$. (b) Same as in (a) but for the mirror-even gapless mode $\boldsymbol{v}$.

and $\delta\boldsymbol{v}$ are coupled and have to be solved together self-consistently. The results of such a calculation are presented in Figs. 3(c,d) of the main text, which displays the spectrum of the phonon modes as a function of the parameter $\gamma D_0/V$.

### D. Lateral stacking shifts

Besides by an external displacement field, the mirror symmetry can also be broken by a lateral stacking shift of an outer layer, which has a strong influence on electronic properties [16]. The top layer's lateral stacking shift breaks not only the mirror symmetry but, unlike an applied external displacement field, also $C_{2z}\Theta$ and $C_3$ symmetries. To describe this shift, the adhesion potential is modified to

$$\mathcal{V}_{\text{ad}} = V \sum_{\nu=1}^{3} \Bigg\{ \cos\left[ \mathbf{G}_\nu \cdot \boldsymbol{r} + \mathbf{b}_\nu \cdot \left( \mathbf{s}^{(2)} - \mathbf{s}^{(1)} \right) \right]$$
$$+ \cos\left[ \mathbf{G}_\nu \cdot \boldsymbol{r} + \mathbf{b}_\nu \cdot \left( \mathbf{s}^{(2)} - \mathbf{s}^{(3)} - \boldsymbol{d} \right) \right] \Bigg\},$$
$$= 2V \sum_{\nu=1}^{3} \cos\left[ \mathbf{G}_\nu \cdot \boldsymbol{r} - \frac{\mathbf{b}_\nu}{2} \cdot (\boldsymbol{v} + \boldsymbol{d}) \right] \cos\left[ \frac{\mathbf{b}_\nu}{2} \cdot (\boldsymbol{u} + \boldsymbol{d}) \right],$$
$$(S38)$$

where $\boldsymbol{d} = d_0 \boldsymbol{a}_\nu$ with $\boldsymbol{a}_\nu$ being a lattice vector of the (rotated) top graphene sheet. With these conventions $d_0 = 1$ corresponds to the top bilayer being in AA stacking and the bottom bilayer being in AB stacking ("A-twist-B" stacking), which has been found to be structurally unstable. Here, we analyze small stacking shifts, $d_0 < 1$, which can lead to metastable structural configurations. More precisely, we use the modified adhesion potential in Eq. (S38) at fixed $d_0$ and determine the textures $\boldsymbol{u}^{(0)}$ and $\boldsymbol{v}^{(0)}$ as above. Note that this procedure can, in principle, lead to $\boldsymbol{u}^{(0)}$ and $\boldsymbol{v}^{(0)}$ that simply have a constant shift of $-\boldsymbol{d}$, which would bring us back to the same phason and shear modes as in the mirror-symmetric case for $d_0 = 0$. However, with small stacking shifts of up to $d_0 \leq 0.50$, we reach distinct metastable states [as shown, e.g., in Figs. 3(e-g) in the main text for $d_0 = 0.25$].

### SIII. ELECTRONIC SYMMETRY BREAKING AND CHIRAL PHONONS

Among the 11 candidate orders analyzed in Ref. 17 for TTG around charge neutrality ($\nu = 0$), there are only two states—denoted by SLP$_\pm$ in that reference—which break $C_{2z}\Theta$ (and any combination of it with U(1) valley symmetry and/or spin rotations) and thus, allow for $L_{\boldsymbol{q}}^z \neq 0$ for

generic $\boldsymbol{q}$. Focusing for simplicity on the TBG-like bands [17] of TTG and denoting Pauli matrices in valley, spin, and band space by $\eta_j$, $s_j$, and $\sigma_j$, the order parameters of the SLP$_+$ and SLP$_-$ states read as $s_0\eta_3\sigma_2$ and $s_0\eta_0\sigma_2$, respectively. As $\Theta = \eta_1\mathcal{K}$, with complex conjugation $\mathcal{K}$, and $C_{2z} = \eta_1$, we see that SLP$_+$ (SLP$_-$) is even (odd) under $\Theta$ and odd (even) under $C_{2z}$, and is therefore characterized by finite loop currents—with nonzero average over the moiré unit cell—of opposite (the same) chirality in the two valleys. We note, however, that $\nu$ close to charge neutrality is not the ideal filling range to observe finite-angular-momentum phonons: SLP$_\pm$ cannot benefit from the intervalley Hund's interaction, and are thus expected [17] to be dominated by the SSLP$_\pm$ states, with order parameters $s_3\eta_3\sigma_2$ and $s_3\eta_0\sigma_2$. However, both of these order parameters are invariant under $is_1C_{2z}\Theta$, wherefore they cannot induce angular momentum in the phonon bands.

The situation is different in the regime $2 \le \nu \lesssim 3$ where Ref. 17 finds spin polarization on top of these candidate orders. In this case, SLP$_\pm$ and SSLP$_\pm$ become equivalent, can induce finite angular momentum in the phonon bands, and are favored energetically beyond a critical value of the displacement field $D_0$. Note further that this doping range is also more interesting as it includes the main superconducting regime of TTG's phase diagram, both close to the magic angle and in the small-twist-angle regime [18, 19].

Next, we first derive a general expression for the angular momentum in terms of the relative displacement fields. Thereafter, we provide further details on the case of broken $\Theta$ and preserved $C_{2z}$ symmetry (SLP$_-$) and then discuss the SLP$_+$ where $C_{2z}$ is broken but $\Theta$ is preserved.

## A. Phonon angular momentum

The total phonon angular momentum of the moiré system is

$$L^z = \rho \int d\boldsymbol{r} \sum_{l=1}^{3} \left( \delta \boldsymbol{s}^{(l)} \times \delta \dot{\boldsymbol{s}}^{(l)} \right)_z = \frac{\rho}{6} \int d\boldsymbol{r} \left[ 3 \left( \delta \boldsymbol{u} \times \delta \dot{\boldsymbol{u}} \right)_z + \left( \delta \boldsymbol{v} \times \delta \dot{\boldsymbol{v}} \right)_z + 2 \left( \delta \boldsymbol{w} \times \delta \dot{\boldsymbol{w}} \right)_z \right]. \tag{S39}$$

If all three modes are decoupled, then we can define the angular momentum $\mathcal{L}^z$ for each individual mode separately, e.g., for mode $\boldsymbol{u}$,

$$\mathcal{L}_{\boldsymbol{u}}^z = \frac{\rho}{2} \int d\boldsymbol{r} \left( \delta \boldsymbol{u} \times \delta \dot{\boldsymbol{u}} \right)_z. \tag{S40}$$

To compute the angular momentum for mode $\boldsymbol{u}$, we need to solve the analogous secular equations when time-reversal symmetry [see Eq. (S55)] or rotational symmetry [see Eq. (S24)] is broken. Our analysis will closely follow Ref. 20. Recall that in deriving the secular equations, we have defined the Fourier transform of $\delta \boldsymbol{u}(\boldsymbol{r}, t)$ as

$$\delta \boldsymbol{u}(\boldsymbol{r}, t) = \int \frac{d\omega}{2\pi} e^{-i\omega t} \sum_{\boldsymbol{q} \in \text{MBZ}} \sum_{\boldsymbol{G}} \delta \boldsymbol{u}_{\boldsymbol{q}+\boldsymbol{G}}(\omega) e^{i(\boldsymbol{q}+\boldsymbol{G})\cdot\boldsymbol{r}}. \tag{S41}$$

Then, the displacement field in second quantization can be written as

$$\delta \boldsymbol{u}(\boldsymbol{r}, t) = \sum_{q=(\boldsymbol{q},\sigma)} \sum_{\boldsymbol{G}} \epsilon_{\boldsymbol{q}+\boldsymbol{G},\sigma} e^{i(\boldsymbol{r}(\boldsymbol{q}+\boldsymbol{G})-\omega_q t)} \sqrt{\frac{\hbar}{2\omega_q m}} a_q + \epsilon_{\boldsymbol{q}+\boldsymbol{G},\sigma}^* e^{-i(\boldsymbol{r}(\boldsymbol{q}+\boldsymbol{G})-\omega_q t)} \sqrt{\frac{\hbar}{2\omega_q m}} a_q^\dagger, \tag{S42}$$

where $q = (\boldsymbol{q}, \sigma)$ includes both the wavevector $\boldsymbol{q}$ in the moiré Brillouin zone (MBZ) and the band index $\sigma$, and $\epsilon_{\boldsymbol{q}+\boldsymbol{G},\sigma}$ is the orthonormal eigenvector ($\epsilon_{\boldsymbol{q}+\boldsymbol{G},\sigma}^\dagger \epsilon_{\boldsymbol{q}+\boldsymbol{G},\sigma'} = \delta_{\sigma,\sigma'}$) that is associated to $\boldsymbol{G}$ in the secular equations. So, the angular momentum is

$$\mathcal{L}_{\boldsymbol{u}}^z = \frac{\rho}{2} \int d\boldsymbol{r} \, \delta \boldsymbol{u}^T i \mathcal{M} \delta \dot{\boldsymbol{u}}, \quad \mathcal{M} = \sigma^y, \tag{S43}$$

$$= \frac{1}{2V} \frac{\hbar}{2} \int d\boldsymbol{r} \sum_{q\,q'} \left[ \epsilon_{\boldsymbol{q}+\boldsymbol{G},\sigma}^\dagger \mathcal{M} \epsilon_{\boldsymbol{q}'+\boldsymbol{G}',\sigma'} \sqrt{\frac{\omega_{q'}}{\omega_q}} a_q^\dagger a_{q'} - \epsilon_{\boldsymbol{q}'+\boldsymbol{G}',\sigma'}^T \mathcal{M} \epsilon_{\boldsymbol{q}+\boldsymbol{G},\sigma}^* \sqrt{\frac{\omega_q}{\omega_{q'}}} a_q^\dagger a_{q'} \right] e^{i(\boldsymbol{q}'+\boldsymbol{G}'-\boldsymbol{q}-\boldsymbol{G})\boldsymbol{r}} e^{i(\omega_q-\omega_{q'})t}. \tag{S44}$$

The integration over space gives $\int d\boldsymbol{r} \, e^{i(\boldsymbol{q}'+\boldsymbol{G}'-\boldsymbol{q}-\boldsymbol{G})\boldsymbol{r}} = V\delta(\boldsymbol{q}'+\boldsymbol{G}'-\boldsymbol{q}-\boldsymbol{G})$, and since $\boldsymbol{q}$ lies in the MBZ, we must have $\delta(\boldsymbol{q}'+\boldsymbol{G}'-\boldsymbol{q}-\boldsymbol{G}) = \delta_{\boldsymbol{q}',\boldsymbol{q}}\delta_{\boldsymbol{G}',\boldsymbol{G}}$. Also using the fact that $\epsilon_q^\dagger M \epsilon_{q'} = -(\epsilon_{q'}^T M \epsilon_q^*)^T$ as well as the commutation

relation $[a_{q,\sigma}, a_{q',\sigma'}] = \delta_{q,q'} = \boldsymbol{\delta}_{q,q'}\delta_{\sigma,\sigma'}$, we arrive at

$$\mathcal{L}_{\boldsymbol{u}}^z = \frac{\hbar}{4V} \int d\boldsymbol{r} \left[ \sum_{q\,q'} \epsilon_{\boldsymbol{q}+\boldsymbol{G},\sigma}^\dagger \mathcal{M}\, \epsilon_{\boldsymbol{q}'+\boldsymbol{G}',\sigma'} \left( \sqrt{\frac{\omega_{q'}}{\omega_q}} + \sqrt{\frac{\omega_q}{\omega_{q'}}} \right) a_q^\dagger a_{q'} e^{i(\omega_q - \omega_{q'})t} \delta_{q,q'} + \sum_q \epsilon_{\boldsymbol{q}+\boldsymbol{G},\sigma}^\dagger \mathcal{M}\, \epsilon_{\boldsymbol{q}+\boldsymbol{G},\sigma} \right],$$
$$= \frac{\hbar}{2} \sum_{\boldsymbol{q},\sigma} \sum_{\boldsymbol{G}} \epsilon_{\boldsymbol{q}+\boldsymbol{G},\sigma}^\dagger \mathcal{M} \epsilon_{\boldsymbol{q}+\boldsymbol{G},\sigma} \left( f(\omega_q) + \frac{1}{2} \right), \tag{S45}$$

where $f(\omega_q) = 1/(e^{\hbar\omega_q/k_B T} + 1)$ is the Bose-Einstein distribution at temperature $T$. The derivations for the modes $\boldsymbol{v}$ and $\boldsymbol{w}$ follow similarly, so the final formula for the total angular momentum is

$$L^z = \frac{1}{6} \left[ 3 \sum_{\boldsymbol{q},\sigma} \ell_{\boldsymbol{q},\sigma}^{\boldsymbol{u}} + \sum_{\boldsymbol{q},\sigma} \ell_{\boldsymbol{q},\sigma}^{\boldsymbol{v}} + 2 \sum_{\boldsymbol{q},\sigma} \ell_{\boldsymbol{q},\sigma}^{\boldsymbol{w}} \right] \left( f(\omega_q) + \frac{1}{2} \right) \equiv \frac{1}{6} \left[ \sum_{\boldsymbol{q},\sigma} L_{\boldsymbol{q},\sigma}^{\boldsymbol{u}} + \sum_{\boldsymbol{q},\sigma} L_{\boldsymbol{q},\sigma}^{\boldsymbol{v}} + \sum_{\boldsymbol{q},\sigma} L_{\boldsymbol{q},\sigma}^{\boldsymbol{w}} \right] \left( f(\omega_q) + \frac{1}{2} \right), \tag{S46a}$$

$$\ell_{\boldsymbol{q},\sigma}^{\boldsymbol{u}} = \hbar \sum_{\boldsymbol{G}} \epsilon_{\boldsymbol{q}+\boldsymbol{G},\sigma}^\dagger \mathcal{M} \epsilon_{\boldsymbol{q}+\boldsymbol{G},\sigma}, \quad \ell_{\boldsymbol{q},\sigma}^{\boldsymbol{v}} = \hbar \sum_{\boldsymbol{G}} \xi_{\boldsymbol{q}+\boldsymbol{G},\sigma}^\dagger \mathcal{M} \xi_{\boldsymbol{q}+\boldsymbol{G},\sigma}, \quad \ell_{\boldsymbol{q},\sigma}^{\boldsymbol{w}} = \hbar \sum_{\boldsymbol{G}} \zeta_{\boldsymbol{q}+\boldsymbol{G},\sigma}^\dagger \mathcal{M} \zeta_{\boldsymbol{q}+\boldsymbol{G},\sigma}, \tag{S46b}$$

where $\xi_{\boldsymbol{q},\sigma}$ ($\zeta_{\boldsymbol{q},\sigma}$) is the orthonormal eigenvector for the $\delta\boldsymbol{v}$ ($\delta\boldsymbol{w}$) modes.

### B. Broken time-reversal symmetry

As discussed in the main text, the broken time-reversal symmetry of the SLP$_-$ state effectively manifests as a Hall viscosity ($F_{\rm HV}$) contribution to the free energy. The viscosity tensor $\eta_{(ij)(kl)}$ is defined as the coefficient of a term with a single time derivative

$$F = \frac{1}{2} \int d\boldsymbol{r} \sum_{l=1}^3 \eta_{(ij)(kl)} s_{ij}^{(l)} \dot{s}_{kl}^{(l)}. \tag{S47}$$

The pair indices $(ij)$ and $(kl)$ are symmetric under exchange $\eta_{ijkl} = \eta_{jikl} = \eta_{ijlk}$ because $s_{ij}^{(l)} = \frac{1}{2}(\partial_i \boldsymbol{s}_j^{(l)} + \partial_j \boldsymbol{s}_i^{(l)})$ is the symmetrized strain tensor. However, the indices are pairwise antisymmetric as $\eta_{ijkl} = -\eta_{klij}$. Due to the antisymmetry, there are only three independent component in 2D: $xxxy, xyyy, xxyy$. Furthermore, with $C_6$ symmetry, there is only one independent coefficient $\eta \equiv \eta_{xyyy}$, which we refer to as the *Hall viscosity* [21, 22]. The other component is related as $\eta_{xxxy} = -\eta_{xyyy}$ and $\eta_{xxyy} = 0$ vanishes. So, the Hall viscosity term relevant

to our system reads

$$F_{\rm HV} = \frac{\eta}{2} \int d\boldsymbol{r} \sum_{l=1}^3 [s_{xy}^{(l)} (\dot{s}_{xx}^{(l)} - \dot{s}_{yy}^{(l)}) - (s_{xx}^{(l)} - s_{yy}^{(l)}) \dot{s}_{xy}^{(l)}]. \tag{S48}$$

Recasting this in terms of the mirror-odd shear mode $\boldsymbol{u}$, the mirror-even phason $\boldsymbol{v}$, and the acoustic phonon $\boldsymbol{w}$ defined in Sec. SI, its contribution to the free energy is

$$F_{\rm HV} = \frac{\eta}{2} \int d\boldsymbol{r} \frac{1}{2} [u_{xy}(\dot{u}_{xx} - \dot{u}_{yy}) - (u_{xx} - u_{yy})\dot{u}_{xy}],$$
$$+ \frac{1}{6}[v_{xy}(\dot{v}_{xx} - \dot{v}_{yy}) - (v_{xx} - v_{yy})\dot{v}_{xy}],$$
$$+ \frac{1}{3}[w_{xy}(\dot{w}_{xx} - \dot{w}_{yy}) - (w_{xx} - w_{yy})\dot{w}_{xy}]. \tag{S49}$$

We now derive how $F_{\rm HV}$ modifies the dynamics of the phonon modes [the secular equations in Eq. (S24)]. Focusing on $\boldsymbol{u}$ and $\boldsymbol{v}$ only, the deviations from the equilibrium solution are defined as $\delta\boldsymbol{u}(\boldsymbol{r},t) = \boldsymbol{u}(\boldsymbol{r},t) - \boldsymbol{u}^{(0)}(\boldsymbol{r})$, $\delta\boldsymbol{v}(\boldsymbol{r},t) = \boldsymbol{v}(\boldsymbol{r},t) - \boldsymbol{v}^{(0)}(\boldsymbol{r})$. The first line of Eq. (S49) then becomes

$$F_{\rm HV} = \frac{\eta}{2} \int d\boldsymbol{r} \frac{1}{2}[(u_{xy}^{(0)} + \delta u_{xy})(\delta\dot{u}_{xx} - \delta\dot{u}_{yy}) - (u_{xx}^{(0)} + \delta u_{xx} - u_{yy}^{(0)} - \delta u_{yy})\delta\dot{u}_{xy}]. \tag{S50}$$

Working in the action description $S[\boldsymbol{u}(\boldsymbol{r},t)] = \int dt\, L[\boldsymbol{u}(\boldsymbol{r},t)]$, we can drop the $u^{(0)}$ terms after integration by parts in $t$ and ignoring the boundary terms. Then, expanding in terms of the deviations leads to

$$F_{\rm HV} = \frac{\eta}{2} \int d\boldsymbol{r} \frac{1}{2}[\frac{1}{2}(\partial_x \delta u_y + \partial_y \delta u_x)(\partial_x \delta\dot{u}_x - \partial_y \delta\dot{u}_y) - (\partial_x \delta u_x - \partial_y \delta u_y)\frac{1}{2}(\partial_x \delta\dot{u}_y + \partial_y \delta\dot{u}_x)],$$
$$= \frac{1}{8}\eta \int d\boldsymbol{r} [(\partial_x \delta u_y + \partial_y \delta u_x)(\partial_x \delta\dot{u}_x) - (\partial_x \delta u_x - \partial_y \delta u_y)(\partial_y \delta\dot{u}_x),$$
$$+ (\partial_x \delta u_y + \partial_y \delta u_x)(-\partial_y \delta\dot{u}_y) - (\partial_x \delta u_x - \partial_y \delta u_y)(\partial_x \delta\dot{u}_y)]. \tag{S51}$$

Integration by part again leads to some cancellations and we end up with a compact form

$$F_{\text{HV}} = \frac{1}{8}\eta \int d\boldsymbol{r}\, [-\partial_i \partial_i \delta u_y]\delta \dot{u}_x + [\partial_i \partial_i \delta u_x]\delta \dot{u}_y.$$  (S52)

Now, the Euler-Lagrange equations are

$$-\frac{\rho}{2}\delta \ddot{u}_i + \frac{1}{8}\eta \epsilon_{ij}\partial_k \partial_k \delta \dot{u}_j + \frac{\lambda+\mu}{2}\partial_i \partial_j \delta u_j + \frac{\mu}{2}\partial_j \partial_j \delta u_i = \delta u_j \left.\frac{\partial^2 \mathcal{V}_{\text{ad}}}{\partial u_i \partial u_j}\right|_{\boldsymbol{u}^{(0)}} + \delta v_j \left.\frac{\partial^2 \mathcal{V}_{\text{ad}}}{\partial u_i \partial v_j}\right|_{\boldsymbol{u}^{(0)},\boldsymbol{v}^{(0)}},$$  (S53)

$$-\frac{\rho}{6}\delta \ddot{v}_i + \frac{1}{24}\eta \epsilon_{ij}\partial_k \partial_k \delta \dot{v}_j + \frac{\lambda+\mu}{6}\partial_i \partial_j \delta v_j + \frac{\mu}{6}\partial_j \partial_j \delta v_i = \delta v_j \left.\frac{\partial^2 \mathcal{V}_{\text{ad}}}{\partial v_i \partial v_j}\right|_{\boldsymbol{v}^{(0)}} + \delta u_j \left.\frac{\partial^2 \mathcal{V}_{\text{ad}}}{\partial v_i \partial u_j}\right|_{\boldsymbol{u}^{(0)},\boldsymbol{v}^{(0)}}.$$  (S54)

After transforming to Fourier space, the only modification to the equations of motion [cf. Eq. (S24)] is the inclusion of a term $\hat{M}$ proportional to $\eta$,

$$\rho \omega^2 \delta \boldsymbol{u}_{\mathbf{G}+\boldsymbol{q}} = \hat{M}_{\mathbf{G}+\boldsymbol{q}}\,\delta \boldsymbol{u}_{\mathbf{G}+\boldsymbol{q}} + \hat{H}_{\mathbf{G}+\boldsymbol{q}}\,\delta \boldsymbol{u}_{\mathbf{G}+\boldsymbol{q}} + 2\sum_{\mathbf{G}'}\sum_{\nu=1}^{3}\left(\mathcal{J}_{\mathbf{G}-\mathbf{G}'}^{\nu}\,\hat{B}\,\delta \boldsymbol{u}_{\mathbf{G}'+\boldsymbol{q}} + \mathcal{K}_{\mathbf{G}-\mathbf{G}'}^{\nu}\,\hat{B}\,\delta \boldsymbol{v}_{\mathbf{G}'+\boldsymbol{q}}\right),$$  (S55a)

$$\rho \omega^2 \delta \boldsymbol{v}_{\mathbf{G}+\boldsymbol{q}} = \hat{M}_{\mathbf{G}+\boldsymbol{q}}\,\delta \boldsymbol{v}_{\mathbf{G}+\boldsymbol{q}} + \frac{1}{3}\hat{H}_{\mathbf{G}+\boldsymbol{q}}\,\delta \boldsymbol{v}_{\mathbf{G}+\boldsymbol{q}} + 6\sum_{\mathbf{G}'}\sum_{\nu=1}^{3}\left(\tilde{\mathcal{J}}_{\mathbf{G}-\mathbf{G}'}^{\nu}\,\hat{B}\,\delta \boldsymbol{v}_{\mathbf{G}'+\boldsymbol{q}} + \mathcal{K}_{\mathbf{G}-\mathbf{G}'}^{\nu}\,\hat{B}\,\delta \boldsymbol{u}_{\mathbf{G}'+\boldsymbol{q}}\right),$$  (S55b)

with $\hat{M}_{\mathbf{G}} = \dfrac{i\omega}{8}\eta \begin{bmatrix} 0 & \mathbf{G}_x^2 + \mathbf{G}_y^2 \\ -(\mathbf{G}_x^2 + \mathbf{G}_y^2) & 0 \end{bmatrix}.$  (S55c)

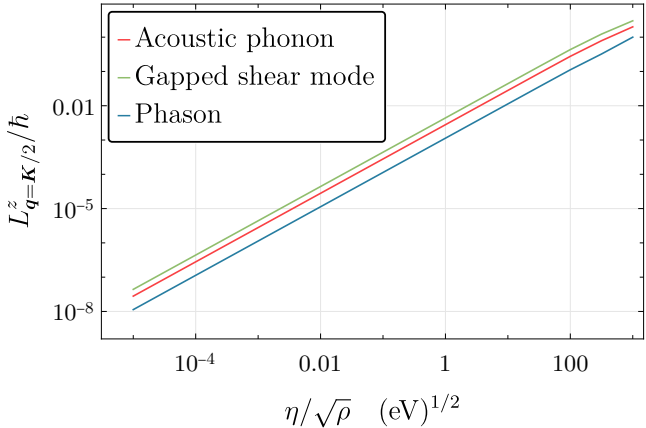

FIG. S4. Variation of the angular momentum of the lowest bands at $\boldsymbol{q} = \boldsymbol{K}/2$ as a function of $\eta$.

Employing this formalism, we iteratively solve for the energies $\omega$ in the presence of a nonzero $\eta$; the associated eigenfunctions are then used to calculate the angular momentum from Eq. (S46a), which is shown in Figs. 4(a–c) in the main text for $\eta/\sqrt{\rho}=1$ $(\text{eV})^{1/2}$. While we treat $\eta$ as a phenomenological parameter of *a priori* unknown magnitude, in Fig. S4, we plot the angular momentum at a generic point in the MBZ for $\eta$ varying over several orders of magnitude. We find that the angular momentum increases, as expected, with the strength of the time-reversal-symmetry breaking, captured by $\eta$.

## C. Broken $C_{2z}$ rotational symmetry

A different possibility for obtaining chiral phonons is to preserve time-reversal symmetry and break $C_{2z}$ symmetry, which is spontaneously broken in the electronic SLP$_+$ phase [17]. In this scenario, time-reversal symmetry implies $L_{\boldsymbol{q}}^z = -L_{-\boldsymbol{q}}^z$, so the total angular momentum $L^z = \sum_{\boldsymbol{q}} L_{\boldsymbol{q}}^z = 0$. However, the angular momentum associated with an individual mode $L_{\boldsymbol{q}}^z$ need not be zero. When $L_{\boldsymbol{K}}^z \neq 0$ (where $\boldsymbol{K}$ is a corner of the Brillouin zone), such chiral valley phonons can lead to interesting selection rules in electron scatterings [23, 24]. Leaving such effects to future work, we focus here on the change in the adhesion potential, which manifests as an additional phase $\phi_p$ [25],

$$\mathcal{V}_{\text{ad}} = V \sum_{p=\pm}\sum_{\nu=1}^{3} \cos\left[\frac{\mathbf{b}_\nu}{2}\cdot(\boldsymbol{v}+p\boldsymbol{u}) - \mathbf{G}_\nu\cdot\boldsymbol{r} + \phi_p\right],$$  (S56)

where $\mathcal{V}_{\text{ad}}$ is invariant under the residual $C_{3z}$ symmetry for an arbitrary phase $\phi_p$ but breaks $C_{2z}$ symmetry if $\phi_p \neq 2n\pi/3$ for $n \in \mathbb{Z}$. With the modification in the adhesion energy, the secular equations in Eq. (S24) are only affected through the Fourier coefficients $\mathcal{J}_{\mathbf{G}}^\nu, \tilde{\mathcal{J}}_{\mathbf{G}}^\nu, \mathcal{K}_{\mathbf{G}}^\nu$. Here, we consider the case where $\phi_p = \phi$, independent of $p$. While Figs. 4(d,e) in the main text illustrated the resultant angular momentum for the specific case of $\phi = \pi/6$, Fig. S5 sketches the variation of the angular momentum with $\phi$ at a generic point in the MBZ; we observe that the angular momentum vanishes at $\phi = \pi/3$, where the $C_{2z}$ symmetry is restored.

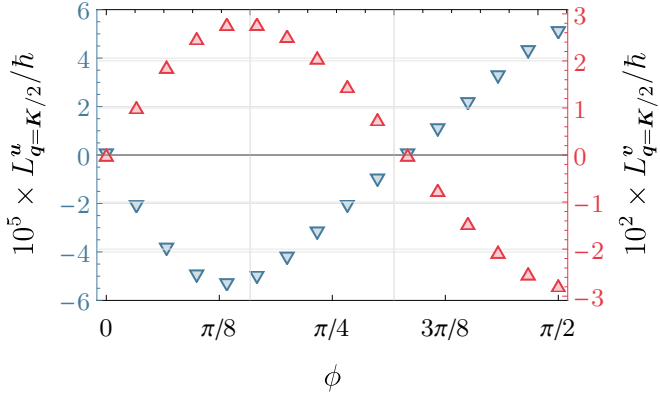

FIG. S5. Angular momentum of the lowest bands at $\boldsymbol{q} = \boldsymbol{K}/2$ as a function of $\phi$. Note that the acoustic phonon does not carry any angular momentum, i.e., $L_{\boldsymbol{q}}^{\boldsymbol{w}} = 0 \,\forall\, \boldsymbol{q}$ in this case.



\end{document}